\documentclass[12pt]{article}
\pdfoutput=1
\usepackage{amsmath}
\usepackage{graphicx,psfrag,epsf}
\usepackage{enumerate}
\usepackage{natbib}
\usepackage{textcomp}
\usepackage[hyphens]{url} 
\usepackage{hyperref}

\newcommand{\blind}{0}

\providecommand{\tightlist}{%
  \setlength{\itemsep}{0pt}\setlength{\parskip}{0pt}}

\usepackage{sectsty}
\usepackage{titlesec}

\usepackage{fullpage}
\usepackage{blindtext}

\usepackage{booktabs}
\usepackage{longtable}
\usepackage{array}
\usepackage{amssymb}
\usepackage{multirow}
\usepackage{wrapfig}
\usepackage{float}
\usepackage{colortbl}
\usepackage{pdflscape}
\usepackage{tabu}
\usepackage{threeparttable}
\usepackage{threeparttablex}
\usepackage[normalem]{ulem}
\usepackage{makecell}
\usepackage[dvipsnames]{xcolor}
\usepackage[font=small,skip=0pt]{caption}
\usepackage{booktabs}
\usepackage{longtable}
\usepackage{array}
\usepackage{multirow}
\usepackage{wrapfig}
\usepackage{float}
\usepackage{colortbl}
\usepackage{pdflscape}
\usepackage{tabu}
\usepackage{threeparttable}
\usepackage{threeparttablex}
\usepackage[normalem]{ulem}
\usepackage{makecell}
\usepackage{xcolor}

\usepackage{amsthm,enumitem}

\begin{document}

\def\spacingset#1{\renewcommand{\baselinestretch}%
{#1}\small\normalsize} \spacingset{1}



\if0\blind
{
  \title{\bf Adaptive local VAR for dynamic economic policy uncertainty spillover}

  \author{
        Niels Gillmann \\
    ifo Institute Dresden and Technische Universität Dresden\\
     and \\     Ostap Okhrin \\
    Chair of Econometrics and Statistics, Technische Universität Dresden\\
      }
  \maketitle
} \fi

\if1\blind
{
  \bigskip
  \bigskip
  \bigskip
  \begin{center}
    {\LARGE\bf Adaptive local VAR for dynamic economic policy uncertainty spillover}
  \end{center}
  \medskip
} \fi

\bigskip
\begin{abstract}
The availability of data on economic uncertainty sparked a lot of interest in models that can timely quantify episodes of international spillovers of uncertainty. This challenging task involves trading off estimation accuracy for more timely quantification. This paper develops a local vector autoregressive model (VAR) that allows for adaptive estimation of the time-varying multivariate dependency. Under local, we mean that for each point in time, we simultaneously estimate the longest interval on which the model is constant with the model parameters.

The simulation study shows that the model can handle one or multiple sudden breaks as well as a smooth break in the data. The empirical application is done using monthly Economic Policy Uncertainty data. The local model highlights that the empirical data primarily consists of long homogeneous episodes, interrupted by a small number of heterogeneous ones, that correspond to crises. Based on this observation, we create a crisis index, which reflects the homogeneity of the sample over time. Furthermore, the local model shows superiority against the rolling window estimation.
\end{abstract}

\noindent%
\vfill

\newpage

\hypertarget{introduction}{%
\section{Introduction}\label{introduction}}
Currently, there is a series of events with international consequences which increase economic uncertainty around the globe. They are the onset of the corona pandemic in early 2020, followed by the attack of Russia on Ukraine in 2022. \cite{Diebold2014} stressed the importance of a timely quantification of the international spillover of current events. In line with this background, our research question is as follows:

``How can we measure \textit{international spillovers} of \textit{current events} in a \textit{timely manner}?''

To measure \textit{current events}, we rely on a dataset created by \cite{Baker2016} with monthly data on Economic Policy Uncertainty (EPU), which is shown to react to current events. More recently, EPU has also been used by the European Central Bank to quantify the increasing uncertainty around the corona pandemic, see \cite{Gieseck2020}. There are many other options for measuring the uncertainty that current events create, see \citet{Bloom2014} for an excellent overview. To mention a few: \citet{Bachmann2013} based on the disagreement of firm survey participants, \citet{Jo2017} on forecast errors of professional forecasters, \citet{Jurado2015} on forecast error variance in a large set of financial and economic variables, and \citet{Creal2017} on interest rate data. Despite the many alternatives, EPU is well suited for our research question since it becomes available with a publication lag of just one month which is much faster than data from official statistics. Additionally, it covers not only developed countries but also countries, where other data is more difficult to obtain. This is also why we prefer EPU to stock market data which also have little publication lag and react to current events but are only available in relatively wealthy countries with an established stock market.

Various researchers have already used EPU data to quantify uncertainty spillovers across countries. One of the first examples is \citet{Kloessner2014}, which employs a spillover method for six countries. They find that spillovers account for one-fourth of variation in EPU across countries and that spillovers change over time. But most papers investigating spillovers, focus on the average spillovers over time instead of the dynamics and quantification of spillovers from recent events, e.g.~\citet{Clausen2019}, \citet{Liow2018}, \citet{Luk2020}, \citet{Tzika2021}. With our research question, we aim to fill the gap in the literature and investigate the temporal dynamics of spillover.

For quantifying \textit{international spillovers}, we use an updated version of the DY-spillover method also used in \citet{Kloessner2014}. This method, created by \cite{Diebold2014}, is based on Forecast Error Variance Decompositions (FEVD) of a Vector Autoregressive (VAR) model. It is widely used to quantify spillovers (see, for example, \citet{Demirer2018} and \citet{Dungey2019}), and has several advantages that make it particularly appealing for our research question. First, it provides an intuitive way to measure spillovers by linking spillover to the question: ``How much of country $A$'s future uncertainty is due to the current situation in country $B$?''. Second, it avoids additional theoretical knowledge for estimation and offers a simple quantitative measurement of spillovers. Third, the framework can easily be adopted into a dynamic setup. Despite the alternative ways of measuring spillovers, such as \citet{Barigozzi2019}, \citet{Engle2012} and \citet{Tobias2016}, the DY-spillover method is the preferred method for our application for its simplicity and interpretability.

Usually, a dynamic setup is adopted through rolling window estimation. However, rolling window sizes are often chosen subjectively and can easily drive the results. Furthermore, too short windows result in large variance, while too long ones result in large bias. Therefore, we propose a data-driven approach to identify meaningful window sizes for dynamic estimation.

Our approach for quantifying international spillovers in a \textit{timely manner} is based on the literature on local parametric estimation, introduced by \citet{Cizek2009} and \citet{Spokoiny2009} through local univariate parametric time series models. Later methodological contributions include \citet{Chen2014a}, \citet{Spokoiny2013}. The local estimation has been applied successfully to many topics, including temperature risk (\citet{Hardle2016b}), crop yields (\citet{Shen2018}), financial risk management (\citet{Fengler2016}), electricity price (\citet{Chen2017a}), and financial (\citet{Hardle2015}) forecasting.

We adapted the framework by \citet{Cizek2009} to the multivariate time series context using finite-sample likelihood ratio tests to test for homogeneous intervals. The works most related to us, which apply the likelihood testing procedure to univariate time series to identify local intervals instead of change points, are \citet{Chen2010}, \citet{Niu2017}, and \citet{Chen2013a}. The local estimation approach is particularly well suited to answer our research question since it is a natural extension of the already established fixed rolling windows. Additionally, this approach allows us to obtain estimation results for our sample's most recent data points, thereby allowing for timely quantification of the spillover of current events.

Our contribution is threefold: 1. We extend the local parametric estimation approach from a univariate autoregressive to a VAR setting. 2. We confirm the successful extension by a set of Monte Carlo simulations by testing if local VAR models can identify homogeneous intervals correctly, even in the presence of structural breaks of various types. 3. In the empirical application to the measurement of EPU spillover, it turns out that spillover of EPU is homogeneous over long episodes, interrupted only by a few major crises, which are the GFC, the European debt crisis, and the trade war. These findings are used to create a \emph{crisis indicator}, highlighting when the sample becomes heterogeneous. Furthermore, total dynamic spillover estimated locally tends to be the same as that estimated by a big rolling window. Only during times of crisis the local approach results in total dynamic spillover that corresponds to a small rolling window size.

The paper proceeds as follows: chapter \ref{C2} introduces the EPU data, chapter \ref{C3} describes the spillover measure, chapter \ref{C4} outlines the local estimation procedure in the VAR context, chapter \ref{C5} contains an extensive simulation exercise, chapter \ref{C6} illustrates the empirical application to EPU, and chapter \ref{C7} concludes. Some specific simulation scenarios are put in the Appendix. 


\section{Economic Policy Uncertainty}\label{C2}
To capture recent events and their potential connections across countries, we use monthly EPU data, \citet{Baker2016}, freely available at \verb|policyuncertainty.com|. The EPU indices are based on newspaper articles classified as related to EPU if they contain specific economic, policy, and uncertainty keywords. The methodology has been used to measure uncertainty in many countries like Australia (\citet{Moore2017}), Chile (\citet{Cerda2018}), and Sweden (\citet{Armelius2017}). Researchers employed this variable to measure uncertainty from current events and estimate its impact on the economy (\citet{Ghirelli2021}, \citet{Pruser2020b}). EPU is also commonly used to investigate spillover effects across countries (\citet{Caggiano2020}, \citet{Nilavongse2021}, \citet{Stockhammar2016}). By now, indices for more than 23 countries are available, with further countries continuously added by groups of researchers who followed the \citet{Baker2016} methodology.

\begin{figure}[ht]
\begin{center}
\includegraphics[width=0.6\textwidth]{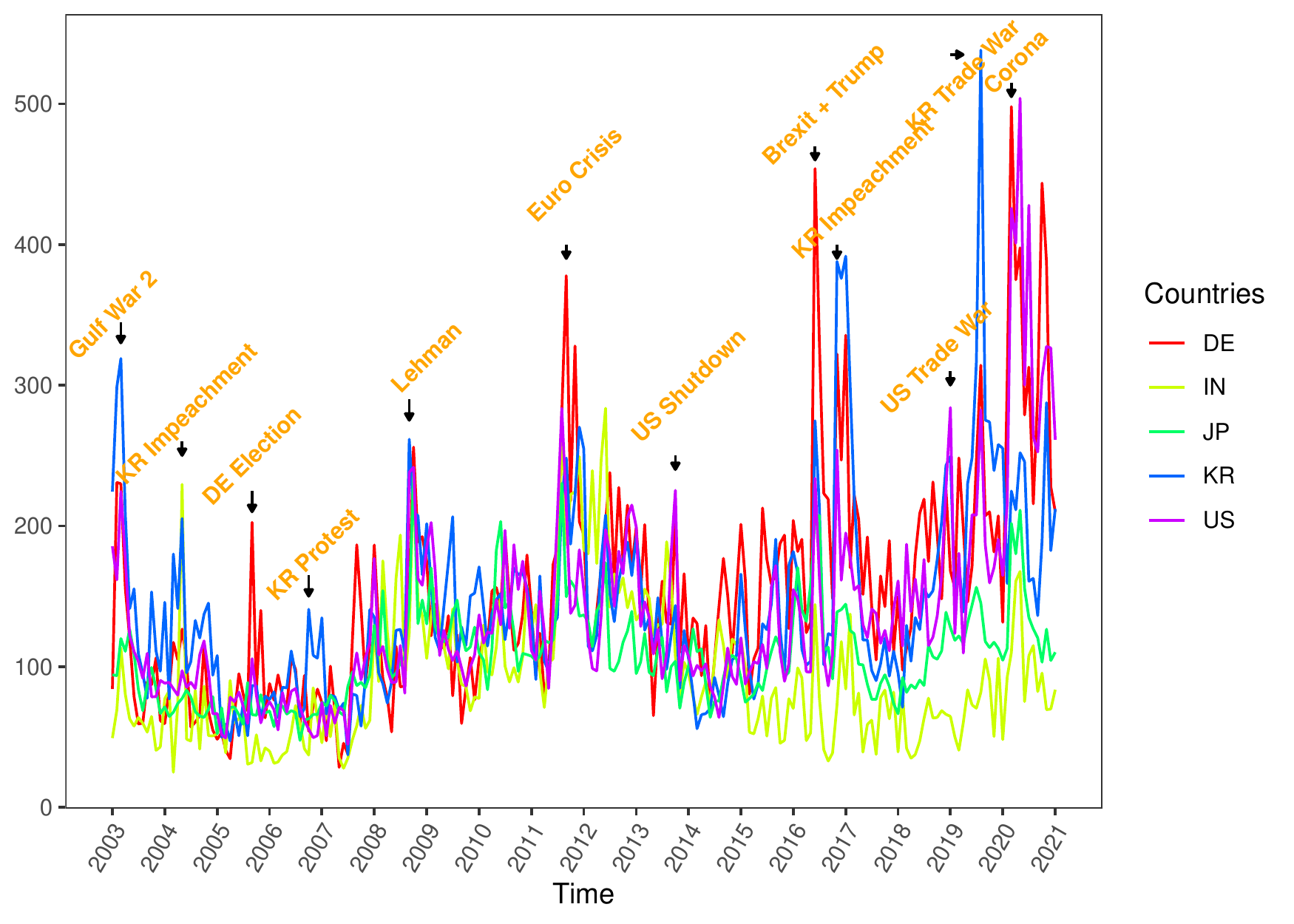}
\end{center}
\caption{Monthly data time series of EPU for Germany (DE), India (IN), Japan (JP), Korean Republic (KR), and the United States of America (US).}
\label{fig:fig1}
\footnotesize
\end{figure}

We chose a set of indices that are based on at least two newspapers from five countries: Germany (DE), India (IN), Japan (JP), South Korea (KR), and the United States of America (US). This constellation is a good mix of large economies from the developed and developing worlds. All chosen countries come from the original database by \citet{Baker2016}. The time-series for the countries in the final dataset are plotted in Figure \ref{fig:fig1}, and some descriptive statistics are shown in Table \ref{tab:tab2SS}. The ADF tests in the table indicate that four out of five series are not stationary when using twelve lags but, became stationary with just one lag. This is a powerful indication that the time series should be modeled locally. Based on data availability, the selected time frame is from $2003M01$ until $2021M01$. 
\begin{table}
\centering
\begin{scriptsize}
\begin{tabular}[t]{lrrrrrrrrrrl}
\toprule
& Med & SD & Min & Max & Skew & Kurt & ADF\textsubscript{12} & ADF\textsubscript{1} & KPSS & $N$ & Time\\
\midrule
DE & 138.53 & 80.91 & 28.43 & 498.06 & 2.60 & 9.51 & 0.22 & 0.01 & 0.10 & 2 & 1993M1-2021M1\\
IN & 80.10 & 50.27 & 24.94 & 283.69 & -1.49 & 5.82 & 0.42 & 0.01 & 0.01 & 7 & 2003M1-2021M1\\
JP & 105.02 & 34.52 & 48.37 & 240.24 & -1.50 & 7.51 & 0.28 & 0.01 & 0.08 & 6 & 1990M1-2021M1\\
KR & 129.70 & 70.87 & 37.31 & 538.18 & 2.74 & 11.68 & 0.05 & 0.01 & 0.10 & 3 & 1990M1-2021M1\\
US & 116.25 & 69.98 & 44.78 & 503.96 & 2.56 & 11.23 & 0.89 & 0.01 & 0.09 & 10 & 1985M1-2021M1\\
\bottomrule
\end{tabular}
\end{scriptsize}
\caption{Summary statistics of the EPU variables, with $p$-values for the Augmented Dickey-Fuller (ADF) and the Kwiatkowski-Phillips-Schmidt-Shin (KPSS) tests. $N$ is the number of newspapers used for construction.}\label{tab:tab2SS}
\end{table}

The period is shaped by major events which resulted in high uncertainty in most countries. There was the second Gulf War ($2003M03$) at the beginning, the Lehman Brothers bankruptcy ($2008M09$) and the Eurozone crisis ($2011M06$) in the middle, and Brexit ($2016M06$) as well as the Trade War ($2019M07$) towards the end. COVID-19 causes the highest spikes at the very end of the sample. The question of time variation in uncertainty transmission is not straightforward. \citet{Angelini2019} estimate a threshold model based on macroeconomic volatility as a proxy for uncertainty which finds strongly increasing impacts of uncertainty during recessions. \citet{Caggiano2020} find similar results using EPU. \citet{Pruser2020a} estimating a TVP model, however, finds that uncertainty transmission is more or less stable over time. With the proposed local model, we are trying to provide more evidence on the time variation of uncertainty, especially since threshold and TVP models require assumptions about the time dynamics while local models do not.

Regarding the general, static transmission of uncertainty, it can be expected that it easily and quickly spills over between countries (\citet{Belke2019}), especially between those with close trade links (\citet{Balli2017}). Furthermore, larger economies will likely transmit uncertainty to smaller economies, as stated by \citet{Tzika2021}. This is partly due to the fact revealed by \citet{Bloom2017} that small open economies have a large probability of being affected by uncertainty shocks of a foreign origin. Also, developing countries were found by \citet{Carriere-Swallow2013} to be more affected by external uncertainty. Therefore, we can expect, for example, the spillover between Germany and the US to be large since they have strong trade links. At the same time, India will probably not receive and transmit much uncertainty to the other countries in the sample as its ties to the other sample countries are weaker. Also, the US, the biggest economy in the sample, will most likely be the primary transmitter of uncertainty. As a small open economy, Korea is expected to experience a lot of spillover from abroad. Just from Figure \ref{fig:fig1}, it can already be seen that Germany, Korea, and the US comove a lot, Japan and India do so to a lesser extent. This observation is also supported by the negative skewness and lower kurtosis of the latter two in Table \ref{tab:tab2SS}.

Applied to the local model, we expect that EPU will be homogeneous for most of the covered period resulting in long intervals of homogeneity. Around the events, as mentioned earlier, the intervals of homogeneity might, however, be relatively short since these events result in sudden surges of EPU. All in all, the dynamics the data display seem to be a good test for the local estimation approach.

\section{Measuring spillover}\label{C3}

Spillover is measured using a framework called connectedness by \citet{Diebold2014}. The framework is prevalent and has been used in many papers, e.g., \citet{Demirer2018}, \citet{Dungey2019}, and \citet{Hale2019}. We are aware of alternative methods to measure spillover and numerous suggestions for improving connectedness, such as \citet{Barunik2018}, \citet{Buse2019}, \citet{Lanne2016}. Since the method is only used to evaluate the local estimation algorithm in an empirical application, we decided that using the original methodology is the most suitable as it has been tested and validated the most. 

Consider a series of $d$-dimensional EPU data $EPU_t = (EPU_{1,t}, ..., EPU_{d,t})^\top$ with $t \in T$ being the time component. It is believed that the size of the share to the $H$-step ahead forecast error of the variance of EPU in the country $i$ due to innovations from EPU in the country $j$, denoted by $C_{ij}(H), i, j = 1, \ldots, N, i\neq j$, can be interpreted as the connectedness between them. Here $i\neq j$ is used to identify relevant variance shares for connectedness. Variance decompositions are used to calculate the variance shares for all countries in the system. The approximating model for the variance decomposition is a VAR model. The traditional VAR relies on orthogonal innovations, whereas connectedness means correlated innovations. Therefore, we used the generalized VAR framework by \citet{Pesaran1998} that allows for correlated innovations by considering the observed distribution of the errors.

Let the cross-sectional dependency between the elements of the EPU vectors be modeled with a VAR as $EPU_t = \sum_{s=1}^{p} \phi_s EPU_{t-s} + \varepsilon_t$, where $\varepsilon \sim N(0, \Sigma)$. When the VAR fulfills the requirements for stationarity, with all roots lying in the unit circle and having a modulus smaller than one, it can be inverted to a Vector Moving-Average (VMA) of infinite order as $EPU_t = \sum_{u=0}^{\infty} A_u \varepsilon_{t-u}$, where the $A_u$'s are $(d \times d)$ moving average coefficient matrices which obey the recursion $A_u=\phi_1 A_{u-1} + \ldots + \phi_p A_{u-p}$, $A_0=\mathfrak{I}$ (the identity matrix), and $A_u=0$ for $u<0$. Based on this setup, country $j$'s contribution to the country $i$'s $H$-step ahead generalized forecast error variance, $C_{ij}(H)$, can be calculated using the previously defined MA coefficients and covariance matrix. The following formula results on the generalized formulation by \citet{Pesaran1998}
\begin{equation}  \label{eq:1} 
  C_{ij}(H) = \frac{\sigma_{jj}^{-1} \sum_{h=0}^{H-1} (e^\top_i A_h \Sigma e_j)^2}{\sum_{h=0}^{H-1}(e^\top_i A_h \Sigma A_h^\top e_j)}, 
\end{equation} 
where $\sigma_{jj}$ is the standard deviation of the error term for the $j$th equation in the VMA, and $e_i$ is a selection vector with zeros except for the $i$'th entry. Due to correlated innovations, the forecast error variance contributions might not add to unity. Following \citet{Diebold2012} each element is divided by its corresponding row sums $\tilde{C}_{ij}(H) = \frac{C_{ij}(H)}{\sum_{j=1}^{N} C_{ij}(H)}$. Having all $\tilde{C}_{ij}(H)$'s, we obtain a spillover table, with diagonal elements representing variance shares and off-diagonal ones the cross-variance shares. Based on the spillovers $\tilde{C}_{ij}(H)$, we can compute total spillover, which is obtained by the total share of the cross-variances in the system $S(H) = \frac{\sum_{i=1, i\neq j}^{N} \sum_{j=1, j\neq i}^{N}\tilde{C}_{ij}(H)}{\sum_{i,j=1}^{N}\tilde{C}_{ij}(H)} \times 100$.

\citet{Diebold2014} further developed the connectedness methodology by considering the variance decompositions as networks. They are more sophisticated than simple networks since the adjacency matrix, which corresponds to the variance decomposition matrix, now contains values ranging between $0$ and $1$ instead of containing either 0 or 1. Additionally, the links are directed so that the link from country $i$ to country $j$ might differ from the link from $j$ to $i$, meaning that the adjacency matrix is not symmetric anymore.

\section{Method} \label{C4}

As mentioned before, this paper aims to measure the impact of current events on international spillover as timely as possible. To quantify spillovers, we use the DY-connectedness measure relying on VAR models. Though rolling window techniques are commonly used in the literature to estimate dynamic spillovers, they are far from optimal. Short windows might work well in the presence of many events associated with high spillover, but they disregard valuable information if there are only a few breaks in the data. Long windows will produce good results in stable times but might incur serious bias when structural breaks are present in the data. Using VAR models with time-varying parameters would require extra assumptions about the dynamics in the data. Hence, we believe that a local estimation approach that allows for time-varying window size is the best solution to measure dynamic spillovers. Additionally, local estimation is quickly applicable since it does not require assumptions on the underlying time dynamics.

\subsection{Time-varying VAR estimation}\label{time-varying-var-estimation}

Let the temporal and cross-sectional dependency of $EPU_t$ be modeled by the VAR process, which differs from the one introduced in Chapter \ref{C3} in the respect that $\phi$ and $\Sigma$ now change with time $t$
\begin{equation} \label{eq:2}
  EPU_t = \phi_{0,t} + \sum_{s=1}^{p} \phi_{s,t} EPU_{t-s} + \varepsilon_t,
\end{equation}
\sloppy
where $\phi_{0,t}$ and $\phi_{s,t} (s=1,\ldots,p)$ are $(d \times d)$ coefficient matrices and the random noise $\varepsilon_t \sim N(0,\Sigma_t)$. To simplify notations, let the parameters be summarised in $\theta_t = (\phi_{0,t}, \phi_{1,t}, \ldots,\phi_{p,t}, \Sigma_t)$. Within VAR models, parameters driving the temporal dynamics are usually assumed to be constants. In practice, however, most processes are not constant over time, and parameters show different behavior during turbulent and calm periods. We deal with this by allowing parameters to vary over time without making specific assumptions about the structure of the time variation. The only assumption we make is that of local homogeneity: For each $\tau \in [1,T]$, there exists a true unknown local interval of homogeneity $I_\tau^* = [\tau - m_\tau^*, \tau]$ over which $\theta_t = \theta$ for $t \in I_\tau^*$. 

The assumption of local intervals represents a good balance between the model's adaptability and its estimation's feasibility. Furthermore, local intervals allow the procedure to handle smooth transitions and sudden jumps of underlying parameters. Thereby we cover both varying coefficients (\citet{Cai2000}) and piecewise constant (\citet{Bai1998}) models. For extensive details on univariate local parametric models and their theoretical properties, we refer to \citet{Spokoiny2009}, and from here on, we closely follow the notation of \citet{Chen2010}. For each local interval $I_\tau$, the local log-likelihood function is defined as
\begin{equation} \label{eq:3}
  \ell(EPU, I_\tau, \theta) = - \frac{m_\tau}{2} \log 2 \pi +  \frac{m_\tau}{2} \log \lvert \Sigma^{-1} \rvert - \frac{1}{2} \sum_{v=\tau-m_\tau +1}^{\tau} \varepsilon_v^\top \Sigma^{-1} \varepsilon_v,
\end{equation}
where $\varepsilon_v = EPU_{v}- \phi_{0,v} - \sum_{s=1}^{p} \phi_{s,v-s} EPU_{v-s}$ and all the parameters are collected in $\theta = (\phi_{0}, \phi_{1}, \ldots, \phi_{p}, \Sigma)$. This results in the following local maximum likelihood (ML) estimator $\tilde{\theta}_\tau = \underset{\theta \in \Theta}{\mathrm{argmax}}\ \ell (I_\tau, \theta)$, with $\Theta$ being the parameter space, where for notational simplicity we denoted $\ell(I_\tau, \theta)$ for $\ell(EPU, I_\tau, \theta)$. 

\subsection{Quality of local estimation}\label{quality-of-local-estimation}

Suppose that for each time point $\tau \in [1, T]$, EPU is driven by a local VAR process with the true (unknown) parameters $\theta^*_{\tau}$ being constant on the homogeneous interval $I_{\tau}^*$. To assess the quality of the local model with parameters $\tilde{\theta}_{\tau}$, we can measure the deviation from the model with optimal parameters $\theta^*_{\tau}$ using a likelihood ratio (LR) statistics as
\begin{equation} \label{eq:4}
  LR(I_{\tau}, \tilde{\theta}_{\tau}, \theta^*_{\tau}) = \ell(I_{\tau}, \tilde{\theta}_{\tau}) - \ell(I_{\tau}, \theta^*_{\tau}).
\end{equation}
There exists a well-established theory for identifying local models with the LR from Equation \eqref{eq:4}; e.g., \citet{Spokoiny2009} and \citet{Cizek2009}. \citet{Polzehl2006} derived a risk bound (RB) which depends on the true parameter $\theta^*_{\tau}$ for the expected deviation \eqref{eq:4} and its $r$th-power transformations with $r > 0$ for an iid sequence of Gaussian innovations
\begin{equation} \label{eq:5}
  E_{{\theta}^*_{\tau}} \lvert LR(I_{\tau}, \tilde{\theta}_{\tau}, \theta^*_{\tau}) \rvert^r \leq RB^r.
\end{equation} 
The introduced bound is nonasymptotic and can be used for any finite interval $I_{\tau}$. It allows us to construct confidence intervals for assessing the quality of estimation, meaning that $\tilde{\theta}_{\tau}$ and the corresponding LR fulfill the risk bound \eqref{eq:5}. Hence, the assessment of the quality of the local estimation is done using the following LR statistics
\begin{equation} \label{eq:6}
  \lvert LR(I_{\tau}, \tilde{\theta}_{\tau}, \theta^*_{\tau}) \rvert^r.
\end{equation}
In practice, the true parameter $\theta^*_{\tau}$ is not known and instead a hypothetical parameter is used for simulating data and calculating the risk bound $RB^r$. Details on the procedure are given in the next section. For a series of models with distributions from exponential families, this risk bound even does not depend on the true parameter, which is, unfortunately, not the case in our model.

\citet{Belomestny2007} show that an optimal choice of an interval of local homogeneity for univariate processes can be obtained via the adaptive procedure. We concentrate on the construction details for a multivariate VAR process in the following. A comprehensive simulation study in the next chapter illustrates the performance of the adaptive procedure in our setting.

\subsection{Adaptive identification of local intervals of homogeneity}\label{adaptive-identification-of-local-intervals-of-homogeneity}

A sequential testing procedure is employed to identify the local homogeneous intervals of the process. Therefore, we consider a finite set of candidate intervals $I_{\tau,k} = \{I_{\tau,1}, \ldots, I_{\tau,K}\}$ with $I_{\tau,k} = [\tau-m_k, \tau]$ and ML estimators $\tilde{\theta}_{\tau}^k$ with $k = 1, \ldots,K$ on each candidate. For each $\tau$, we start with the shortest possible interval $I_{\tau,1} = [\tau-m_1, \tau]$, assumed to be homogeneous. From there on, we extend the interval backward and test whether parameter $\tilde{\theta}_{\tau}^1$ is also well suited on the next bigger interval $I_{\tau, 2} = [\tau - m_2, \tau]$. If the hypothesis is not rejected, we consider $I_{\tau, 2}$ to be homogeneous and continue extending until the largest possible interval is reached.

The procedure is performed using the obtained ML estimators $\tilde{\theta}_{\tau}^k$ and the LR statistic \eqref{eq:6}. The only difference is that now the true $\theta^*_{\tau}$ is replaced with the best-known one -- the adaptive estimator $\hat{\theta}_{\tau}$, which will be determined sequentially at each backward-looking step and formalized below. The adaptive estimators will differ for each $\tau$ since it depends on $I_{\tau}^*$. Furthermore, since $I_{\tau, 1}$ is assumed to be always homogeneous, the procedure starts with $\hat{\theta}_{\tau} = \tilde{\theta}_{\tau}^1$ and then continues either until the last interval $I_{\tau, K}$ or where the test statistic does not exceed the critical value as
\begin{equation} \label{eq:7}
  \lvert{ LR(I_{\tau,k}, \tilde{\theta}^k_{\tau}, \hat{\theta}_{\tau})}\rvert^{r} \leq \zeta^r_k, \quad k = 2,\ldots,K,
\end{equation}
with $\zeta^r_k$ being the critical value at step $k$ and is described in more detail below. The test statistic measures the difference between the current local ML estimator $\tilde{\theta}_{\tau}^k$ and the adaptive estimator $\hat{\theta}_{\tau}$ over a possible $k$-th interval of local homogeneity $I_{\tau,k}$. If the test statistic is small, there is no significant change in the dynamics, and \eqref{eq:7} is not violated. We thus cannot reject the null of local homogeneity and adopt the new estimator $\tilde{\theta}_{\tau}^k$ as the \emph{adaptive estimator} $\hat{\theta}_{\tau} = \tilde{\theta}_{\tau}^k$. Suppose the test statistic is bigger than the critical value. In that case, it indicates that the adaptive estimator $\hat{\theta}_{\tau}$ for the current point in time $\tau$ cannot be extended further backward and is only valid until $k-1$. Therefore, the iterative procedure is terminated and $\tilde{\theta}_{\tau}^{k-1}$ is accepted as the optimal i.e.~the \emph{adaptive estimator} $\hat{\theta}_{\tau}$ for the current
${\tau}$.

When testing the procedure with simulations, we realized that for a few $\tau$, we sometimes obtain implausible interval series like $I_{\tau, 6}, I_{\tau+1, 1}, I_{\tau+2, 6}$, corresponding to the intervals of the lengths $m_{\tau, 6}$, $m_{\tau+1, 1}$, and $m_{\tau+2, 6}$ respectively. For the real data as $EPU_t$, this behavior is hardly empirically interpretable, but formally this is because the procedure takes each $\tau$ individually. Hence, we implemented an additional \verb"if-condition" that ensures that the final intervals $I_\tau$ do not allow for sudden unexpected jumps: If $m_{\tau,k} \neq m_{\tau-1,k}$, then $m_{\tau,k} = m_{\tau,k_{max}}$, where $k_{max}$ is the index of the interval with the largest LR: $ \max_{k \in [1, K]}\{LR(I_{\tau, k}, \tilde{\theta}^{k}_{\tau}, \hat{\theta}_{\tau})\}$. It was furthermore verified in the simulations that this additional restriction does not affect the results but only produces more smooth and better interpretable changes of the identified intervals of homogeneity. 

The whole procedure can be thus summarized as follows
\begin{enumerate}
\def\labelenumi{\arabic{enumi}.}
\tightlist
\item
  initialization: Set $\hat{\theta}_{\tau} = \tilde{\theta}_{\tau}^{1}$ and $k = 2$
\item 
  while $\lvert{ LR(I_{\tau,k}, \tilde{\theta}^{k}_{\tau}, \hat{\theta}_{\tau}) }\rvert^{r} \leq\zeta^r_k$ and $k \leq K$\\
  \hspace*{15mm}do $\hat{\theta}_{\tau} = \tilde{\theta}_{\tau}^{k}$, $k = k + 1$.
\item
  set $\hat{m}_{\tau} = m_{\tau,k-1}$ and $\hat{\theta}_{\tau} = \tilde{\theta}_{\tau}^{k-1}.$
\item
  check if $\hat{m}_{\tau} = m_{{\tau}-1,k}$. If not, $\hat{m}_{\tau} := m_{\tau, k_{max}}$, where $k_{max}$ is the index of the interval with the largest $LR$.
\end{enumerate}

\subsection{Critical values}\label{critical-values}
The critical values $\zeta_k^r$ are obtained through Monte Carlo simulations. There are two main ingredients. The first is the empirical version with the expectation being replaced by the theoretical risk bound $RB^r$ based on $\theta^*_\tau$. The second is the empirical counterpart which is the deviation between the MLE estimator $\tilde{\theta}_\tau^k$ and the adaptive estimator $\hat{\theta}_\tau$ measured by a likelihood ratio. Additionally,$\frac{k}{K}$ as a normalizing factor to make estimates based on different $k$ comparable and $\rho$, a tuning factor, to ensure that the $LR$ and $RB$ match, are needed. 

The whole procedure, which is similar to \citet{Chen2014a}, \citet{Hardle2015} and \citet{Shen2018}, is summarized in the following algorithm, where we use the notation $\ell(X_i, I_{\tau, k}, \tilde\theta^k_\tau)$ to highlight, that the likelihood is evaluated on the interval $I_{\tau, k}$ of the sample $X_i$ using the parameter $\tilde\theta^k_{i,\tau}$:
\begin{enumerate}
\def\labelenumi{\arabic{enumi}.}
\tightlist
\item
 simulate $N = 10^4$ homogenous processes $X_{i,t}$, $i=1, \ldots, N$ from a fixed $\theta^*$.
\item
 use \eqref{eq:5} to calculate $\widehat {RB}^r_k = \frac{1}{N}\sum_{i=1}^N\left|\ell(X_i, I_{\tau, k}, \tilde\theta^k_{i,\tau}) - \ell(X_i, I_{\tau, k}, \theta^*)\right|^r$, for $k = 2, \ldots, K$.
\item
 set initial critical values $\zeta_k^r = \infty$.
\item 
  using \eqref{eq:6} to get $\hat{\theta}_{i, \tau}(\zeta_k^r)$ for each sample $i=1, \ldots, N$ select $\zeta_k^r$ over $k = 2,\ldots,K$ by
  \begin{equation*}
    \zeta_k^r = \arg\min_\zeta\left|\frac{1}{N}\sum_{i=1}^N\left|\ell(X_{i}, I_{\tau, k}, \tilde\theta^k_{i\tau}) - \ell\{X_{i}, I_{\tau, k}, \hat\theta_{i, \tau}(\zeta)\}\right|^r -  \rho \frac{k}{K}\widehat{RB}^r_k\right|.
  \end{equation*}
\end{enumerate}

\subsection{Selection of $\rho$ and $r$}\label{selection-of-rho-and-r}

There are two choices to be made for the calibration of critical values: $\rho$ and $r$. Keeping $r$ fixed while increasing $\rho$ will lead to smaller critical values. On the other side, leaving $\rho$ fixed while increasing $r$ will lead to bigger critical values.

\citet{Hardle2015} suggest $r=0.5$ and $\rho=0.5$ in a univariate setting, while \citet{Chen2018} recommend $r=0.5$ in a functional AR model. Since the selection of $\rho$ is often arbitrary, we follow the idea from \citet{Cizek2009} and determine it by minimizing prediction errors. In detail, we estimate local models over a grid of values for $\rho$ ranging from 0.01 to 1 (including 0.5). From each set of local models, we predicted from the estimated VAR the $\widehat{EPU}_{t, \rho}$ for each $\rho$ from the grid. In the end, we compute the Mean Absolute Percentage Error (MAPE) for each $\rho$ and select the $\rho$ which corresponds to the smallest MAPE as $\hat\rho = \underset{\rho > 0}{\mathrm{argmin}} \frac{1}{T} \sum_{t=1}^{T} \left\lvert \frac{EPU_t - \widehat{EPU}_{t, \rho}}{EPU_t} \right\rvert$. The algorithm results in a potentially different $\rho$ and, therefore, a different critical value for each dataset. This is important because different datasets might need a higher or a lower break detection sensitivity depending on the amount of general heterogeneity. 

\section{Simulation}\label{C5}

We perform a Monte Carlo study to investigate the performance of the local estimation procedure. There are three criteria that the procedure should fulfill. First, it should not detect a break when there is none in the data. Second, it should detect a break, when there is one. Third, it should not recover to the maximum length if there is a second break shortly after the first. Therefore, datasets that include tests for all three conditions are generated. But before results are presented, the choice of parameters and the setup has to be described to ensure reproducibility.

\subsection{Simulation design}\label{simulation-design}

We use a finite set of $K+1 = 7$ candidate intervals based on a geometric grid $m_i=[m_0a^k]$, with $m_0 = 12$ and $a = 1.25$, where the $[x]$ means the largest integer smaller than $x$, which results in the following set of interval lengths: $m_{\tau,k} = \{12,15,19,23,29,37,46\}$. Note that the first interval corresponding to a length of 12 is always assumed to be homogenous and works as a baseline against which to compare the candidate intervals. In our setting, the smallest interval corresponding to twelve months i.e., one year seemed a good tradeoff between timeliness and estimation accuracy. The geometric grid is preferred over the linear grid since it generally yields better results in the simulation exercise and is also used by most of the literature \citep{Cizek2009, Hardle2015, Spokoiny2009}.

Parameters for testing our algorithm are obtained from fitting two-dimensional VAR(1) models to EPU\footnote{We modified the parameters obtained from the data a bit since the original ones resulted in a large break which was easy to detect. The parameters presented here will result in a small break and are, therefore, challenging. Results for higher dimensions are shown in Appendix A1.}. A short lag is sufficient in our setting because our set of intervals is also relatively short. The resulting parameters can be found in Table \ref{tab:tab4}. Two points in the selection of the parameters are worth noting: a) the structural break will contain changes in both $\phi_0$ and $\phi_1$; b) changes in the parameters are small. This is the way the real data typically change. Theoretical studies usually vary just one single coefficient, but we are interested in empirical applications where more than one coefficient might change.

\begin{table}
  \centering
  \begin{tabular}[t]{c|c|c|c} 
    \toprule
    $\theta_1$ 
    & \multicolumn{3}{c}{$\phi_{0} = \left[ \begin{array}{c} 29.00 \\ 132.00 \end{array}\right]$ $\phi_1 = \left[ \begin{array}{cc} 0.71 & 0.08 \\ 0.13 & 0.08  \end{array}\right]$} \\
    \hline
$\theta_2$
    & \multicolumn{3}{c}{$\phi_{0} = \left[ \begin{array}{c} 31.00 \\ 130.00 \end{array}\right]$ $\phi_1 = \left[ \begin{array}{cc} 0.63 & 0.00 \\ 0.12 & 0.23  \end{array}\right]$} \\
    \bottomrule
  \end{tabular}
  \caption{Parameters for the simulation study with $d = 2$ and $p=1$}
  \label{tab:tab4}
\end{table}

The simulation results are divided into three scenarios ranging from ``easy'' to ``difficult''. Here we present only Scenario 1, and the two other scenarios can be found in Appendix A1. Scenario 1 contains a single break: $x_1,\ldots,x_{84} \sim \theta_1, x_{85},\ldots,x_{146} \sim \theta_2$. The simulated data are referred to as $x_t$, and we generate each dataset 250 times. The notation ``$\sim \theta_1$'' means that observations are generated from a VAR(1) with parameter $\theta_1$. In Appendix A2, we present the distribution of test statistics for each scenario and compare it with the corresponding selected critical values. The same exercise is also performed for different sets of parameters to guarantee robustness, with results available upon request. Since the growth of the interval of homogeneity for each $\tau$ goes backward, an initial set of 46 observations corresponding to the longest interval has to be discarded.

For each scenario, we calculate four different results. First is optimal $\rho$ and its restriction. Then, results for two fixed $\rho$'s, namely the most often chosen $\rho$ from the optimal algorithm and a $\rho = 0.5$. Finally, results with the optimal $\rho$ algorithm but without the additional restriction are also shown.

Results are presented as three combined plots to show the input data and the algorithm. The setup is as follows: The plot on the top shows the simulated input data with the break. Its purpose is to serve as an orientation for when something should happen in the local process. The middle panel shows the resulting intervals of homogeneity for each $\tau$. Plotted are means (solid lines) and medians (dashed lines) over 250 repetitions. The plot at the bottom depicts the values of the likelihood ratio tests for the accepted interval plus one for each $\tau$ to show whether the algorithm starts to react after the breaks appear in the input data. Please note that the values might be a combination of tests for different $k$'s since each repetition might stop at a distant $k$. The horizontal red line indicates the critical value for the longest possible interval, $k = 6$. A test value below means that the local interval could have been extended more. There is a dashed line through the top plot that highlights the point when the parameter set for the input data changes.

\subsection{Simulation results}\label{simulation-results}

\begin{figure}[ht]
\begin{center}
\includegraphics[width=0.8\columnwidth]{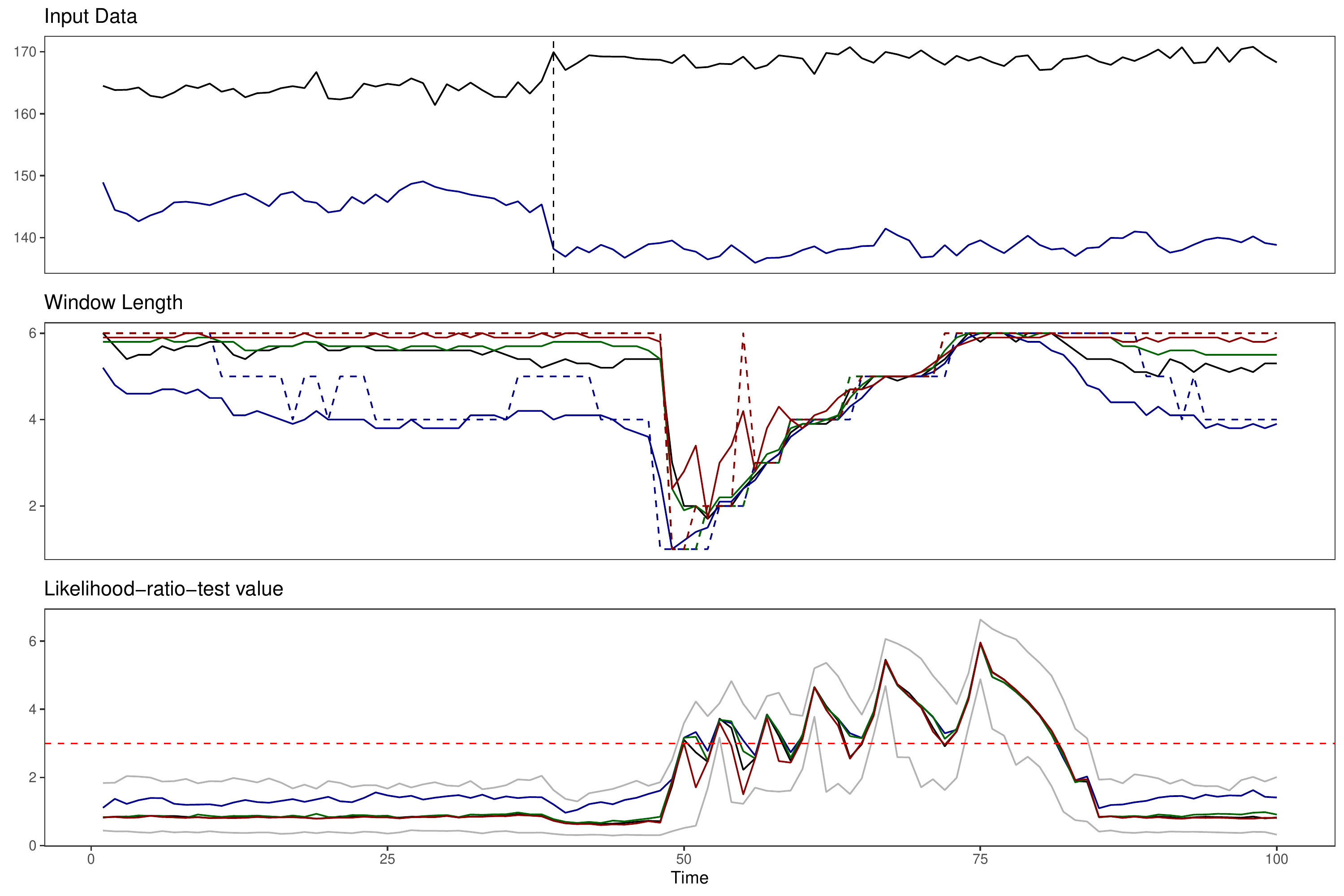}
 \end{center}
\caption{Simulation results - One break. \label{fig:fig2}}
\footnotesize
\textit{Top}: Simulated 2dim VAR. The vertical line indicates a break. \textit{Mid}: Identified Intervals for: optimal $\rho$ (black), $\rho = 0.5$ (blue), $\rho = 0.088$ (green), optimal $\rho$ with no restrictions (red). \textit{Bottom}: Test statistics for $k+1$ with $\zeta_6 = 3.0$ (horizontal red). Bounds for 5$\%$ and 95$\%$ bounds are depicted in grey.
\end{figure}

Figure \ref{fig:fig2} shows the simulation results for Scenario 1 based on 250 repetitions. There is just one break located in the middle of the observation period. It is marked by a dashed line to help visualize exactly where the break occurs. The middle plot shows that all four specifications detect the break. The reaction to the break is that the window length jumps down to one. After the break, it slowly increases to maximum length again. The recovery process is characterized by a step function with six constant regions, which we call \emph{stairs}. These six stairs correspond to the six intervals. So, while the stairs for the lower intervals are shorter, the stairs for the higher intervals are more prolonged, corresponding to the predetermined intervals $I_{\tau, k}, \ k=1,\ldots, K-1$. The number and length of the preselected intervals can determine the number of stairs and their length.

As expected, the procedure with the optimal $\rho$ performs best. While the medians are almost identical, with only the median window length for $\rho = 0.5$ being visibly lower than six, there are some differences in the means. Specification for $\rho = 0.5$ results in an average window length of only four instead of six. On the other side, specification with an optimal $\rho$ and no restrictions does not result in flat steps for lower window lengths and instead displays some spikes, particularly for the first and second steps. The algorithm, without any restrictions, using $\rho = 0.088$, results in the most symmetric distribution of window lengths, as the mean is closest to the median.

The lower means happen because, over the 250 repetitions, each run will have a few $\tau$'s, where there is a false alarm, and the algorithm mistakenly selects a window length of one. Most false alarms appear for specification $\rho = 0.5$, which features the highest value of $\rho$ for this setting. This large value results in a low critical value, so false alarms happen more often when the test statistic exceeds the critical value. In the univariate setting, $\rho = 0.5$ was a good choice. Still, in our multivariate setting with small window sizes, we need higher critical values to prevent false alarms while detecting true breaks.

The upwards spikes in the stairs of specification with optimal $\rho$ with no restrictions happen because here, the algorithm jumps down to a small window length when the break occurs but directly jumps up again in a set of repetitions so that the mean is upwards biased. In the other specifications, this upward bias is prevented by using the additional restriction, which keeps the selected window length low for an adequate time. This is evidence that applying the additional restriction in the multivariate setting with small window sizes makes sense.

The likelihood ratio test values in the bottom plot are virtually identical across specifications, where only $\rho=0.5$ results in slightly higher values. It is reassuring to see that 95\% of test statistics during the homogeneous part of the sample are below the critical value for the longest possible interval. Then, as soon as the break in the input data occurs, there is a slight reaction. However, it takes a bit until the break shows in the window lengths. This is because the test statistic only becomes large once all observations inside the smallest interval come from $\theta_2$.




\section{Empirical Application}\label{C6}

The goal of this chapter is twofold. First, it should provide evidence that empirical data are mostly homogeneous over time since crises are rare events. Second, it should illustrate that longer intervals are more suitable for measuring dynamic spillover because, during homogeneous periods, there is little risk of bias when using longer intervals for estimation. Therefore, at first, the local estimation algorithm identifies the longest homogeneous interval for each $\tau$ in the sample of EPU for Germany, India, Japan, South Korea, and the US. Second, these intervals will be used to calculate the Diebold Yilmaz connectedness measure, a common way to measure spillover, similar to a rolling window approach. The connectedness measure is detailed in Chapter \ref{C3}.  The same set of interval lengths $m_{\tau, k} = \{12, 15, 19, 23, 29, 37, 46\}$ as for the simulation in Chapter \ref{C5} based on a geometric grid is used for the setup of the local estimation algorithm. As a benchmark for the second part, rolling windows with windows of the length $w$ of the size of the shortest ($w = 12$) and longest ($w = 37$) interval are additionally applied. Note that the last window ($w = 46$) cannot be selected because of how the local estimation algorithm is designed. Therefore a length of $w = 37$ constitutes the longest possible window length. Furthermore, since during the estimation of the intervals of homogeneity, intervals extend backward in time from $\tau$, we need to discard the first $46$ observations so that the first $\tau$ is $2006M11$. The last one is then $\tau = 2021M01$.

\subsection{A crisis indicator based on local homogeneity}\label{crisis indicator}

For all possible pairs of EPU in the five selected countries (DE, IN, JP, KR, and US), bivariate VAR models are estimated to identify homogeneous intervals. This strategy is chosen because estimating just one VAR with all five countries would flatten out the breaks, which only affect a fraction of the countries. By estimating bivariate VAR models, heterogenous episodes that only occur in a few country pairs are also considered.

The length of the intervals of homogeneity can be treated as an indicator of structural breaks that happened in the recent past. Moreover, structural breaks in economic policy are rare events and, in history, were mirrored through various crises. Based on this logic, we define the \emph{crisis indicator} CI between countries $i$ and $j$, which reflects the homogeneity of the multivariate process at each time point $\tau$, as
\begin{equation} \label{eq:10}
  CI_{i,j} = 1 - \frac{\hat k_{\tau, i, j} - 1}{K-1},
\end{equation}
where $\hat k_{\tau, i, j}$ is the index of the estimated length of the interval of homogeneity for the VAR(1) model that is modeling EPU for countries $i$ and $j$ and $K-1$ is the index of the largest possible interval ($K=6$ in our paper). The resulting values fall in the interval $[0,1]$, with an indication of the crisis via $CI_{i,j}=1$ indicating the shortest interval length corresponding to $k_{\tau, i, j}=1$, and no crisis with $CI_{i,j}=0$ the longest one corresponding to $k_{\tau, i, j}=K$. Therefore, a crisis indicator of zero means no heterogeneity in the sample and the longest possible interval of homogeneity. As soon as the indicator starts increasing towards one, we obtain shorter intervals of homogeneity, leading to an increase in heterogeneity, indicating that some structural change is happening.

Suppose all the models possess the longest homogeneity interval for the given $\tau$. In this case, the causal effects between EPU for different countries follow a fixed structure over the given period. This implies that no shocks were present during this period in economic policy for \emph{all} the involved countries. If for some country pairs for a given $\tau$, the homogeneity interval is short, this implies a structural break in the economic policy modeling in this pair. Therefore averaging the crisis indicator $CI_{i,j}$ over all the pairs of countries, we obtain the \emph{global crisis indicator} 
\begin{eqnarray*}
  CI &=& \frac{2}{d(d-1)}\sum_{i = 1}^{d - 1}\sum_{j = i + 1}^d CI_{i,j} \\
  &=& \frac{K}{K-1} - \frac{2}{d(d-1)(K-1)}\sum_{i = 1}^{d - 1}\sum_{j = i + 1}^d \hat k_{\tau,i,j},
\end{eqnarray*}
with $d$ being the number of countries. Instead of the mean, one can use a median that neglects the magnitude of the interval indices, thus leading to a more robust specification. 

\begin{figure}[ht]
\begin{center}
\includegraphics[width=0.6\columnwidth]{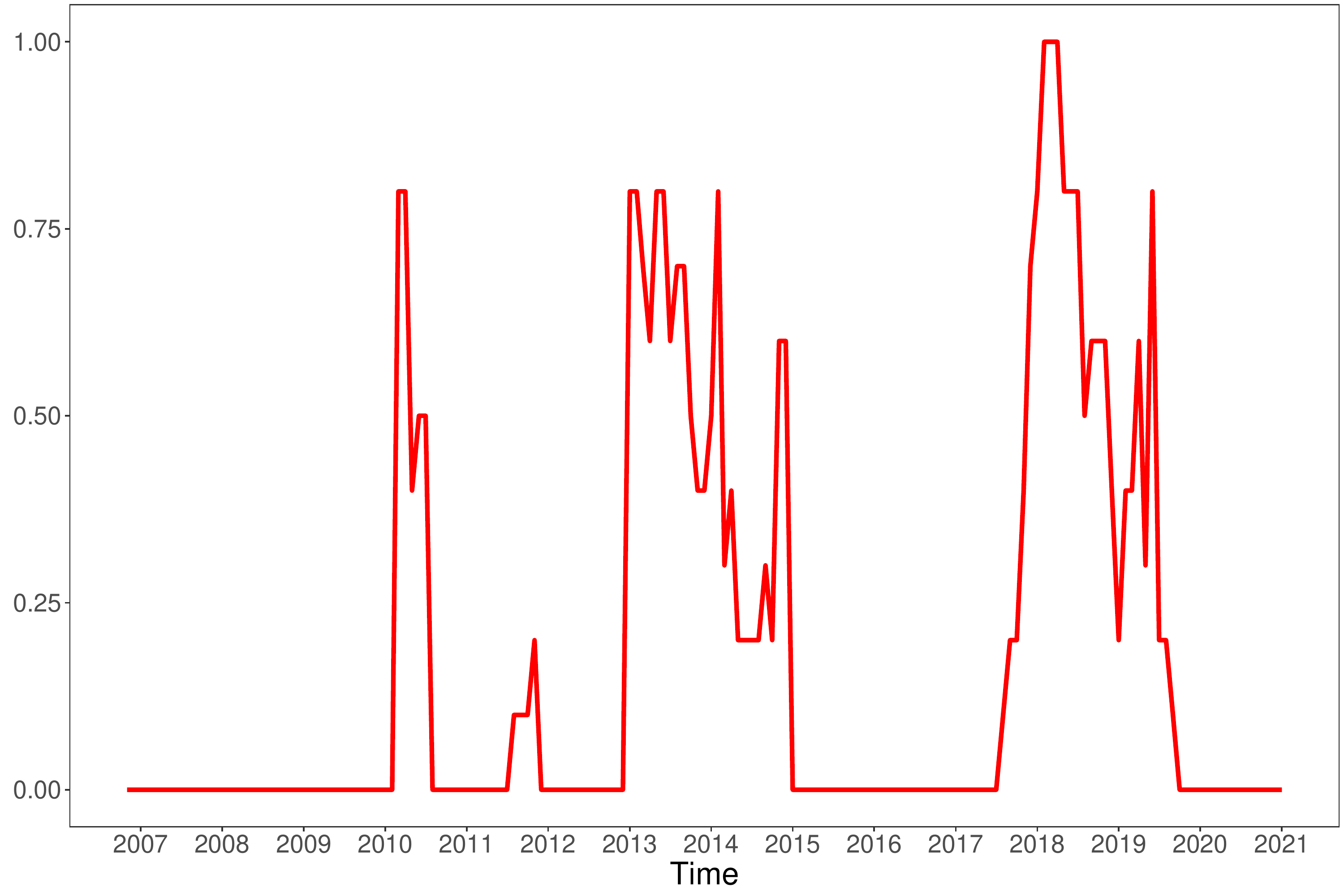}
\end{center}
\caption{A global crisis indicator based on local homogeneity. \label{fig:fig3}}
\footnotesize 
The crisis indicator is obtained by transforming the intervals of local homogeneity ranging from 1 to 6 into a 0,1 Index. A value of 0 indicates homogeneity and, therefore, no crisis. A value of 1 indicates the absence of homogeneity and, thus, a crisis in all pairs. 
\end{figure}

The CI estimated for EPU is presented in Figure \ref{fig:fig3}. Here, the identified intervals of a local homogeneity of all ten pairs based on EPU in the five selected countries are used to compute the global crisis indicator $CI$. It has long periods of homogeneity where no crisis exists in any country pair. These episodes usually span several years. The homogeneous episodes are interrupted by three common crises in nearly all the pairs. The first crisis lasted less than one year and corresponded to the global financial crisis, which resulted in a sudden shock to economic policy. Furthermore, it affected economic policy in countries with less integration into the global financial market, like India, to a lesser degree. Therefore, the global crisis indicator does not reach a value of one during this crisis. The second crisis lasted two years, which is more than double the duration of the first one. It corresponds to the Eurozone debt crisis, a prolonged episode of uncertainty about economic policy in the Eurozone. Simultaneously, there was uncertainty about economic policy in the US due to a debt ceiling debate and the risk of a government shutdown. Both issues involved discussions over a longer period to resolve underlying problems. This is reflected in two years when the global crisis indicator stayed elevated. Still, the crisis indicator again did not reach a value of one because of India and the same reasoning.

The last crisis in the sample is different since it does attain the value of one, meaning that it affected economic policy in all country pairs in the sample. This crisis can be attributed to two simultaneously occurring events: the Brexit referendum and the election of Donald Trump as president of the United States. Both events severely increased the uncertainty about international cooperation, especially regarding trade agreements. Since India has historical links to the UK and trades a lot with the US, economic policy in India was affected by this increase in uncertainty as well, resulting in a value of one for the global crisis indicator. The crisis also lasted for a long time since the involved governments of the UK and the US at this time did not swiftly act to dispel worries about their political agendas. This is in line with earlier warnings from researchers such as \cite{Baker2016}, that governments should communicate policies transparently and predictably since they otherwise risk causing high economic policy uncertainty. Furthermore, \cite{Caggiano2020} found that rising EPU in influential countries such as the US can trigger large spillovers to smaller countries. Putting that together with the finding in \cite{Davis2016} that global economic policy uncertainty peaked during the Brexit referendum further supports our conclusion that Brexit and the election of Donald Trump resulted in the most global crisis of economic policy in our sample.

All in all, the proposed global crisis indicator allows a classification of recent events into periods of crises and non-crises based on the degree of homogeneity that the data exhibit during the respective events. It helps to assess how global the specific crises were and also sheds light on how long each crisis lasted. From the application, economic policy crises last longer if governments do not clarify their agendas or try to dispel worries about future policy goals. We also learned that economic policy in developing countries like India is not that much affected by internal political issues in the developed world or issues regarding the global financial market. However, economic policy in developing countries is affected by uncertainty about international cooperation and trade. Lastly, a minor shortcoming of this setting with monthly frequency data is that our algorithm can only detect breaks with a delay. Using monthly frequency will therefore require some patience to classify recent crises. Daily data might be more suited for monitoring ongoing crises.   

\subsection{Spillover estimation results}\label{spillover-estimation-results}
\begin{figure}[ht]
\begin{center}
\includegraphics[width=0.81\columnwidth]{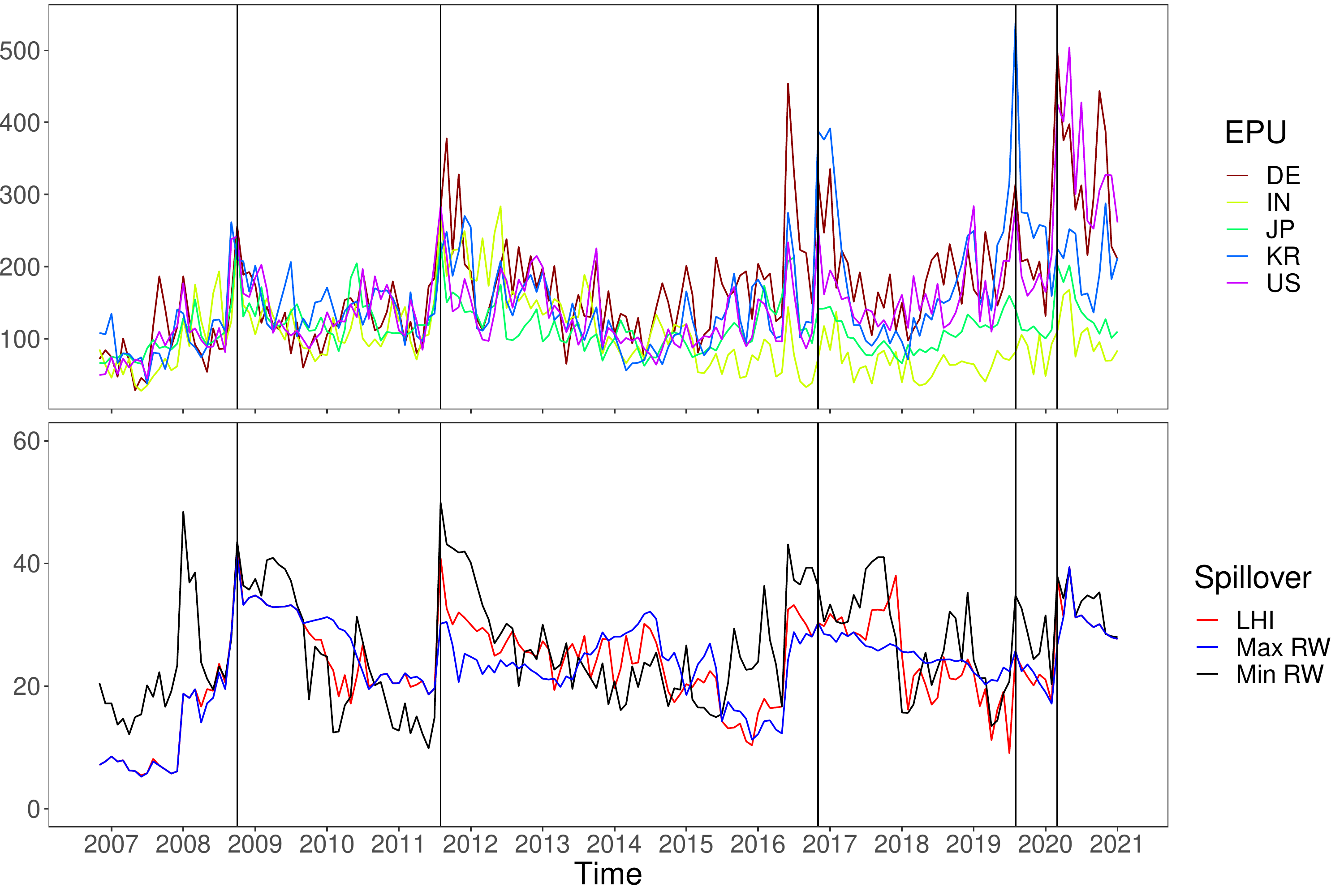}
\end{center}
\caption{Spillovers. \label{fig:fig4}}
\footnotesize 
\textit{Top}: EPU data for the five selected countries. \textit{Bottom}: Total spillover based on all pairs of 2dim VARs. LHI stands for Local Homogeneous Intervals. Min and Max RW, respectively, stand for fixed windows of length 12 and 37 months. 
\end{figure}

Papers like \cite{Angelini2019, Caggiano2020} have shown that uncertainty spillovers from Chapter \ref{C3} vary over time. The intervals of homogeneity identified in the previous section are window lengths of a rolling window procedure selected in a data-driven way. They feature primarily long windows, interrupted by a few sequences of short windows during episodes of crises. This section uses them to estimate bivariate rolling window VARs for calculating DY spillovers. The top panel of Figure \ref{fig:fig4} shows the actual EPU data, and the bottom one shows the spillover for the different window lengths with LHI standing for Local Homogeneous Intervals. In line with most of the literature, uncertainty spillovers vary over time. They increase during high economic policy uncertainty and decrease again when uncertainty is low. This means that long windows do not, as might be worried by some, smoothen out the countercyclicality in the data. Instead, the long windows can adequately capture the data's features. The data has five peaks marked with vertical lines, which correspond to: 1. The global financial crisis in October 2008. 2. The European sovereign debt crisis in August 2011. 3. The US and UK political struggle from June 2016 onwards. 4. The trade war between the US and China, as well as the EU in 2019. 5. The outbreak of COVID-19 in 2020. From the previous section, we know that the third crisis was more global than the first and second one, affecting all countries in the sample. This did not result in higher spillover as the spillover during the two previous crises attained the same level. The third crisis differs from the two other crises, however, in the time persistence of the elevated spillover level. It seems that a crisis being more global will lead to a prolonged episode of high spillover instead of a more pronounced peak. 

During the global financial crisis, spillover jumped up drastically in October 2008. This was driven by the US, where Lehman Brothers declared bankruptcy on September 15th, triggering the crisis. The European debt crisis was a prolonged episode of increased uncertainty. The highest spillover in our sample is recorded in August 2011. Among the countries, it is Germany and Japan who push up spillover this month. For Germany, it is clear that discussions about bailouts for other European countries created uncertainty. Regarding Japan, the government intervened in currency markets to prevent the yen from rising, which would make Japanese exports less competitive on the international market. This intervention also seems to have created significant spillovers of economic policy uncertainty. In June 2016, spillover jumped up. This is, when the first results of the Brexit referendum became public. It caused stock markets around the world to fall. In our sample, the event is most clearly visible in the country pairs of India and Korea, India and the US, as well as Korea and the US. This is probably because especially India and the US are closely linked to the UK. During the trade war, spillover is highest in August 2019 for the country pairs involving the Korean Republic. This is because Korea announced that it would scrap a military information agreement with Japan, which reacted to a Japanese decision to tighten high-tech exports to Korea. This announcement prompted stark criticism from the US and the international community, spreading uncertainty about Korea's future economic policy to the world.
The Covid-19 crisis caused the strongest spillover in March 2020. While spillovers among pairs involving either Japan or the US were highest, pairs with India generally resulted in very low spillovers during this month. Japan introduced mandatory quarantine for travelers from China and Korea in early March. Korea responded by suspending visas for all Japanese citizens traveling to Korea. Furthermore, Japan announced in March 2020 that it would postpone the Olympic Games by one year and implement a range of economic measures to stop the spread of the virus. In the US, Donald Trump tried to downplay concerns about COVID-19 as long as possible. But in March 2020, he was forced to announce a national emergency and had to acknowledge that the US would be heading for a recession. Therefore, the US transmitted uncertainty because of Trump's unclear and hesitant communication. In contrast, Japan transmitted uncertainty due to the large range of measures, of which the impact on economic development was unclear at the time of the announcement.

When comparing the different methods for calculating spillover, one immediately notices that spillovers are much higher when based on small windows. This is because small windows result in increased variance. The high spillover, resulting from small windows, does not reflect changes in the data but the estimation uncertainty of the small windows. The spillover based on the LHI is similar to the spillover based on long windows, which is more stable and has lower variance. The proposed method improves the estimation of spillover compared to rolling windows since it sometimes results in higher peaks, also staying high for longer periods, than the long windows, meaning that our local estimation algorithm reacts more to sudden increases in the data while still maintaining a low variance. This is particularly noticeable around 2011 during the European sovereign debt crisis and in 2016 during the US and the UK political struggle. Here, the long windows smoothen out some variation in EPU, failing to represent time-varying spillover adequately. 

\begin{figure}[ht]
\begin{center}
\includegraphics[width=0.81\columnwidth]{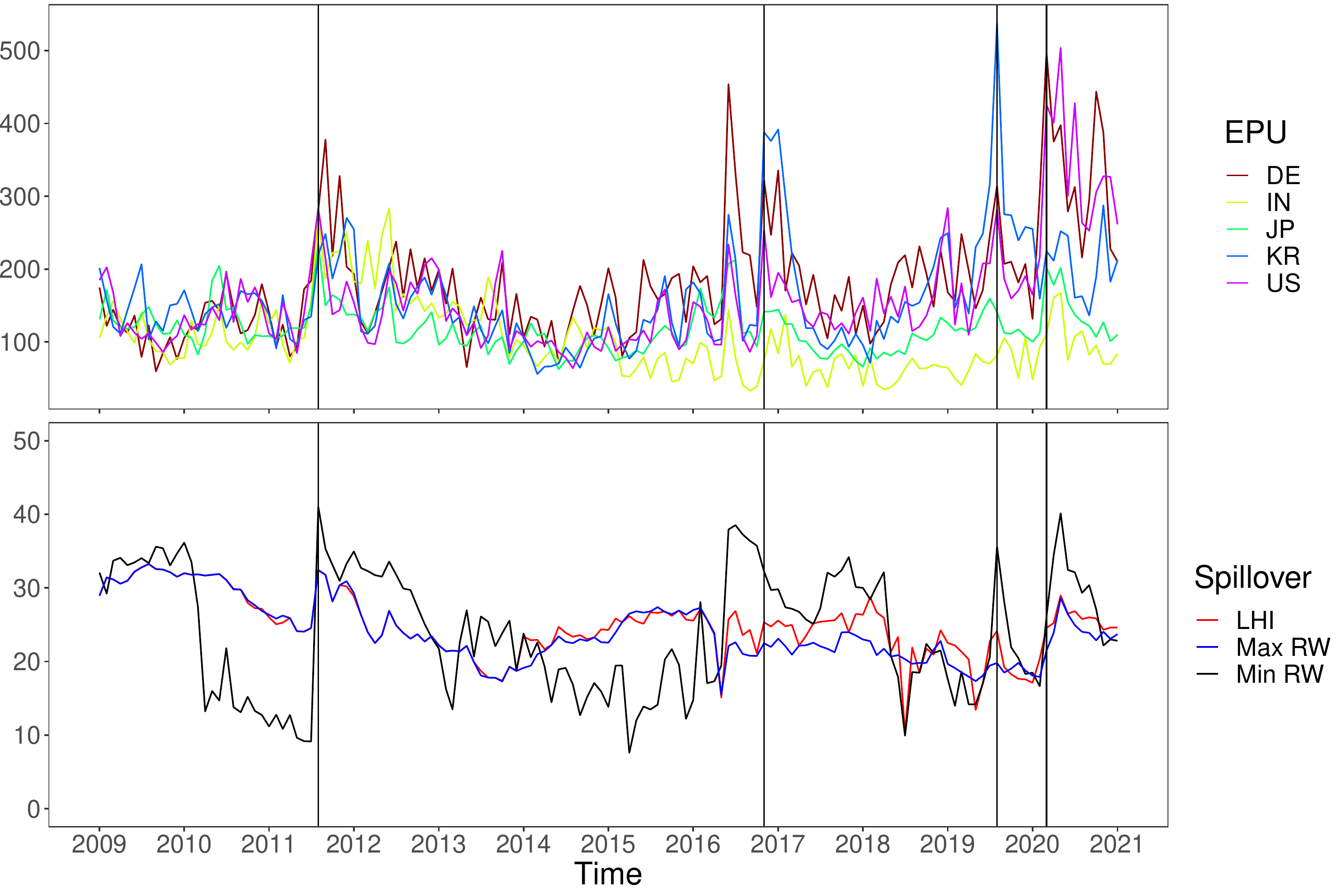}
\end{center}
\caption{Spillovers based on window sizes used in the literature. \label{fig:fig5}}
\footnotesize 
\textit{Top}: EPU data for the five selected countries. \textit{Bottom}: Total spillover based on all pairs of 2dim VARs. LHI stands for Local Homogeneous Intervals. Min and Max RW, respectively, stand for fixed windows of length 18 and 71 months. 
\end{figure}

The window sizes used in the previous estimation represent our choice based on ideal windows for monthly frequency. In empirical applications, researchers used different window sizes. The variance of the window sizes employed in the literature ranges from 18 months in \cite{Yin2014} to 72 months in \cite{Clausen2019}. Therefore, we use a new grid of windows: $m_{\tau,k} = \{18, 23, 29, 36, 45, 57, 72\}$. Since we now have longer intervals, we have to discard more data at the start of the sample so that now the starting month is 2009M01. The results of this exercise are depicted in Figure \ref{fig:fig5}.

The variance between the spillover based on the shortest and longest window increased compared to the previous figure. Especially the spillover estimated with the smallest window size seems very large compared to the two other curves. Interestingly, our algorithm indicates that the longest window is again preferred for spillover calculation since the spillover based on the LHI is again very close to the longest window, even though the longest window is now 71 months instead of just 37 months. The only period were the longest window size would smoothen out too much variation is again around 2016, when the spillover based on our method shows more variation and higher spillover. Based on this finding, one should be careful when selecting window sizes for rolling window estimation not to choose too small windows. Too small windows will result in high values of spillover during crises. They are, therefore, tempting. These results are based more on estimation uncertainty than increased economic policy uncertainty.

\section{Conclusion}\label{C7}

An existing algorithm for detecting locally homogeneous intervals in a univariate setting is adapted to the multivariate VAR context. Through a series of Monte Carlo simulations, the algorithm is shown to perform well in this multivariate context, even when intervals are small and breaks occur within short periods from each other or as smooth changes over time. The algorithm can be applied to many settings by modifying the set of intervals.

In the paper's last chapter, the algorithm is applied to identify homogeneous intervals and estimate the spillover of EPU across countries. It turns out that empirical data are primarily homogeneous, and breaks only occur during a limited number of common crises. This feature is exploited to create a crisis indicator, which measures how the homogeneity of the sample changes over time. The most significant episode of non-homogeneity turned out to be the time around Brexit and the election of Donald Trump, when all countries experienced increased uncertainty about economic policy. Even the global financial crisis had less impact on economic policy uncertainty since it affected developing countries to a lesser degree. From the spillover estimation part, it does not seem necessary to have window sizes that vary at each $\tau$ since breaks occur only during specific periods and for short amounts of time. Most of the time long window sizes are adequate for quantifying spillovers. This means that sophisticated assumptions about time dynamics are unnecessary. Small adaptions to the traditional rolling window approach are sufficient to take care of potential bias when estimating time dynamics.

Avenues for further research are the application of the algorithm to empirical settings with different time frequencies of the data, such as quarterly or weekly data. Furthermore, the crisis indicator is an interesting topic for further research. It might indicate uncertain times earlier than commonly available uncertainty indicators since heterogeneous patterns in the data usually characterize uncertain times.

The authors report there are no competing interests to declare.

\bibliographystyle{Chicago}
\small{
\bibliography{Data/2023_Adaptive-local-VAR_Arxiv.bib}
}

\newpage


\begin{center}
{\LARGE\bf Appendix \\ Adaptive local VAR for dynamic economic policy uncertainty spillover}
\end{center}

\newpage
\spacingset{1.8} 

\appendix
\hypertarget{simulation-robustness}{%
\section{\texorpdfstring{Simulation robustness
\label{A1}}{Simulation robustness}}\label{simulation-robustness}}
This appendix shows simulation results for scenarios two and three as well as results for $d=4$ for all three scenarios. The set of intervals is the same for $d=4$, but we do need to use a different set of parameters to estimate from, and therefore, we also obtain new risk bounds. As in the setting with $d=2$, we simulate three scenarios ranging from easy to complex. The results are presented in the following three sections, one for each scenario.

\subsection{Simulation robustness for Scenario 1: One break}\label{simulation-robustness-for-scenario-1-4d}
\begin{figure}[ht]
\begin{center}
\includegraphics[width=0.7\columnwidth]{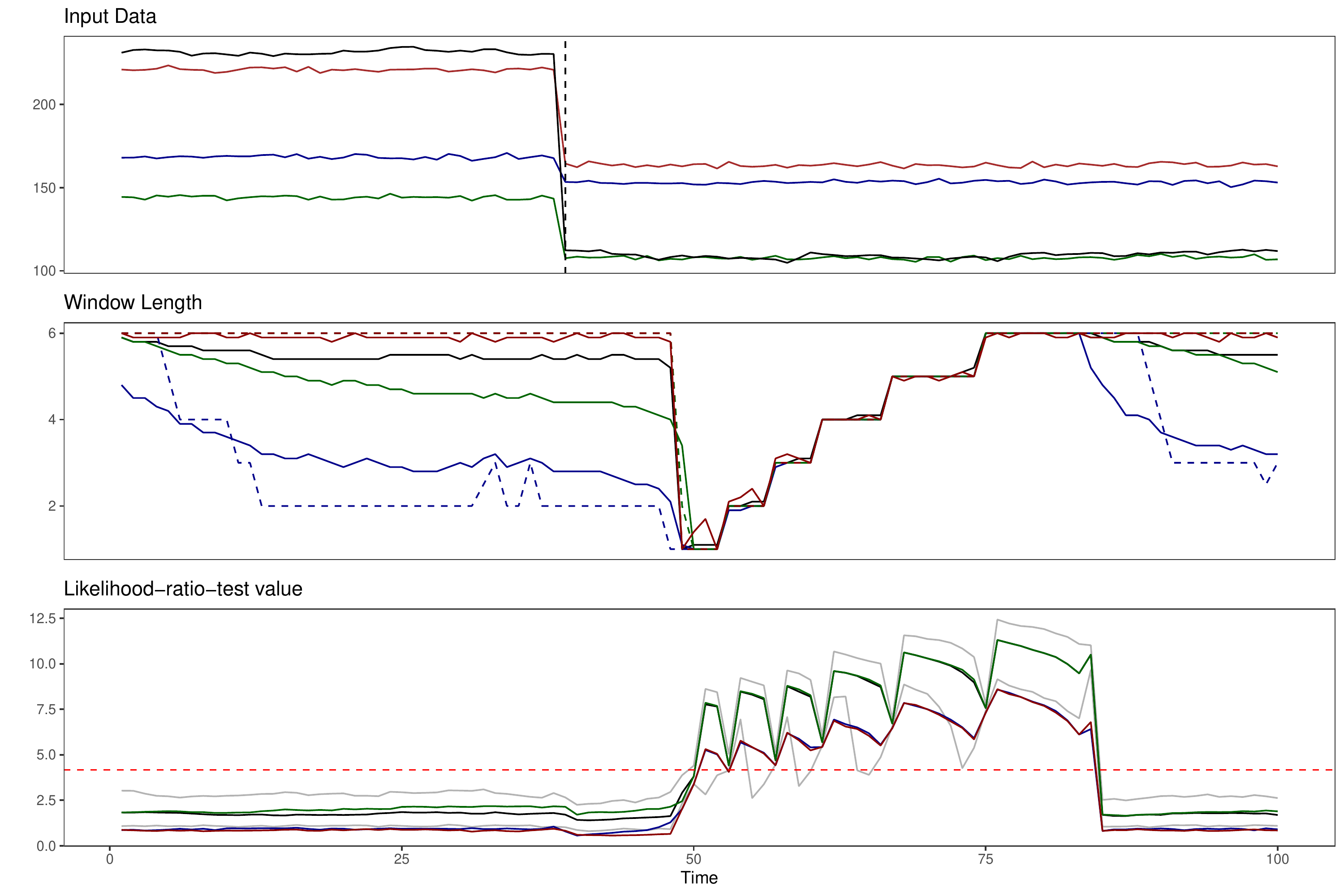}
\end{center}
\caption{Simulation robustness $d=4$ - One break. \label{fig:fig6}}
\footnotesize
\textit{Top}: Simulated 4dim VAR. The vertical line indicates a break. \textit{Mid}: Identified Intervals for: optimal $\rho$ (black), $\rho = 0.5$ (blue), $\rho = 0.088$ (green), optimal $\rho$ with no restrictions (red). \textit{Bottom}: Test statistics for $k+1$ with $\zeta_6 = 4.1$ (horizontal red). Bounds for 5$\%$ and 95$\%$ bounds are depicted in grey.
\end{figure}

This section only contains results for $d=4$, which are depicted in Figure \ref{fig:fig6}. In general, the results are very similar to the case when $d=2$. The only noticeable difference is that there is now a larger heterogeneity in the mean (solid lines) of the different specifications during the homogeneous part of the sample. The medians (dashed lines) are still all perfectly detecting the break.

\subsection{Simulation robustness: Scenario 2 - Two breaks}\label{simulation-robustness-scenario-2}
\begin{figure}[ht]
\begin{center}
\includegraphics[width=0.8\columnwidth]{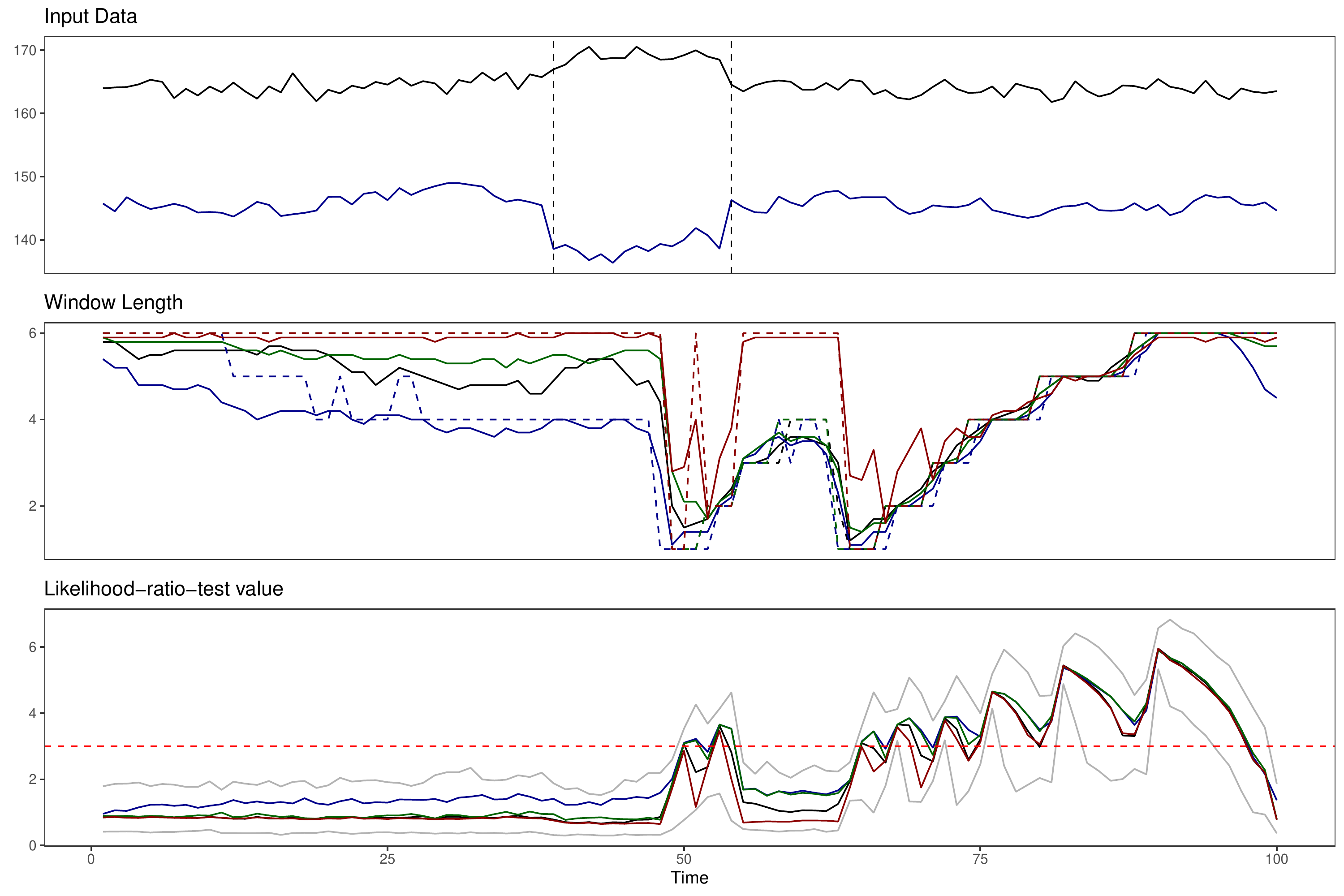}
\end{center}
\caption{Simulation robustness $d=2$ - Two breaks. \label{fig:fig7}}
\footnotesize
\textit{Top}: Simulated 2dim VAR. Vertical lines indicate breaks. \textit{Mid}: Identified Intervals for: optimal $\rho$ (black), $\rho = 0.5$ (blue), $\rho = 0.088$ (green), optimal $\rho$ with no restrictions (red). \textit{Bottom}: Test statistics for $k+1$ with $\zeta_6 = 3.0$ (horizontal red). Bounds for 5$\%$ and 95$\%$ bounds are depicted in grey.
\end{figure}

The results for Scenario two are depicted in Figure \ref{fig:fig7}. It contains two breaks that occur within a brief period: $x_1,\ldots,x_{84} \sim \theta_1, x_{85},\ldots,x_{99} \sim \theta_2, x_{100},\ldots,x_{146} \sim \theta_1$. Since they are barely visible, two dashed lines clarify where the breaks happen. All four specifications detect both breaks. The behavior before the first break is very similar to the first scenario; it, however, changes afterward. Suppose the additional restriction is not applied instead of a small upward bias. In that case, the algorithm returns to the maximum window length before jumping down again when the second break happens. This happens because observations from both sets of parameters mix. Shortly after the first break, most observations in the tested windows again belong to $\theta_1$. Because of this, specification for optimal $\rho$ with no restrictions will result in long homogeneous window lengths furthermore too early.

The means of the other three specifications show an inverted u-shape that only grows to a window length of four before jumping down to one again. The mix of observations from both sets of parameters causes this smoother behavior. Nevertheless, the algorithm can still identify the second proper break.

After the first break, the values of the likelihood ratio test fall below the threshold so that the window length would be the maximum again. The restriction prevents the procedure from jumping up, however. This is another piece of evidence why the additional restriction is beneficial. It keeps results clear and identified breaks visible.

The downward movements at the end of the sample in the middle and bottom panels indicate the return to homogeneity. The test statistic falls below the critical value, and the mean of specification $\rho = 0.5$ returns to a window length of four again.

\begin{figure}[ht]
\begin{center}
\includegraphics[width=0.7\columnwidth]{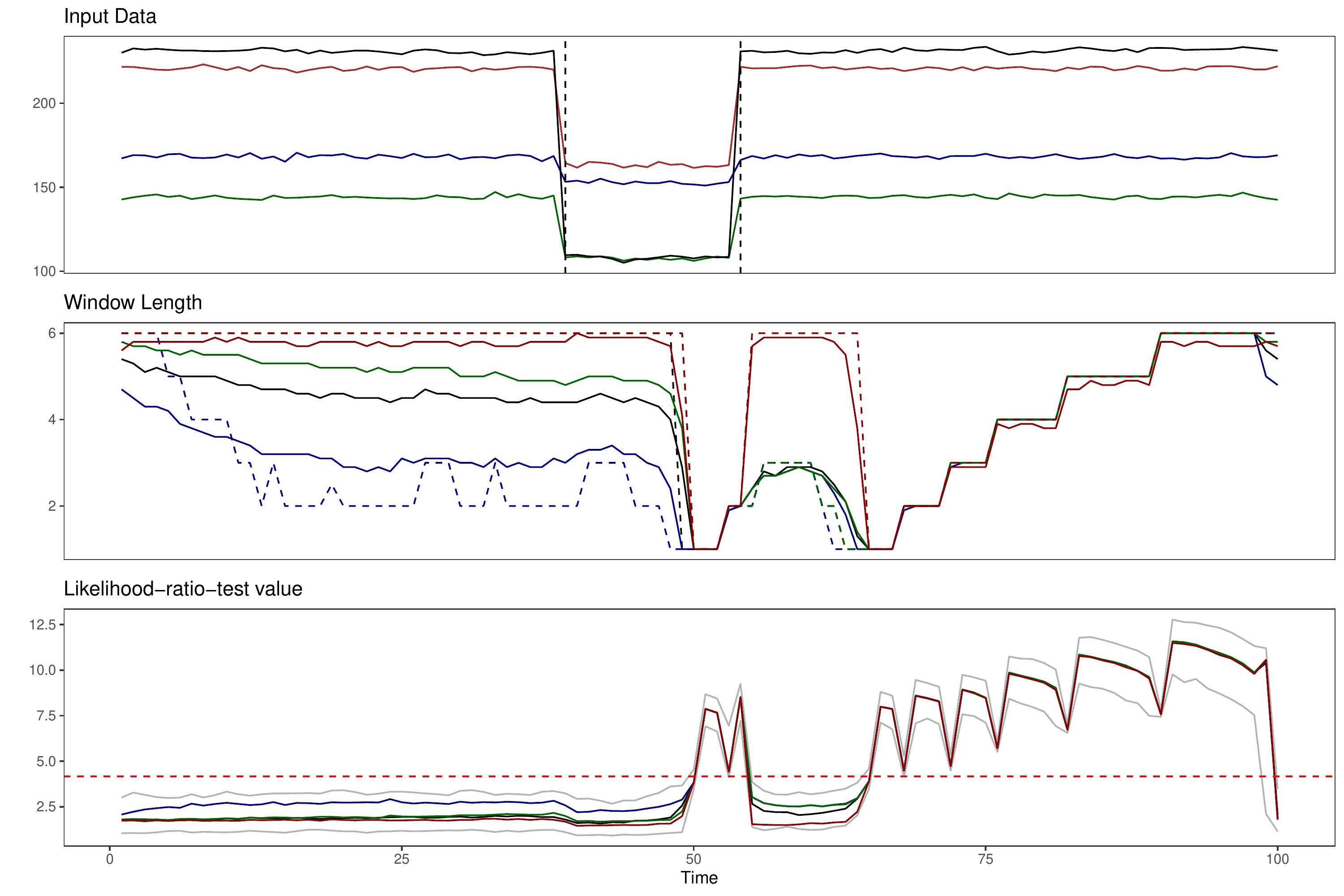}
\end{center}
\caption{Simulation robustness $d=4$ - Two breaks. \label{fig:fig8}}
\footnotesize
\textit{Top}: Simulated 4dim VAR. Vertical lines indicate breaks. \textit{Mid}: Identified Intervals for: statistics $\rho$ (black), $\rho = 0.5$ (blue), $\rho = 0.088$ (green), optimal $\rho$ with no restrictions (red). \textit{Bottom}: Test statistics for $k+1$ with $\zeta_6 = 4.1$ (horizontal red). Bounds for 5$\%$ and 95$\%$ bounds are depicted in grey.
\end{figure}

Results for the second scenario with $d=4$ are depicted in Figure \ref{fig:fig8}. Again, the results are very similar to the case when $d=2$ except for a bit more heterogeneity in the means of the different specifications during the homogeneous part of the sample.

\subsection{Simulation robustness: Scenario 3 - Smooth break}\label{simulation-robustness-scenario-3}
\begin{figure}[ht]
\begin{center}
\includegraphics[width=0.802\columnwidth]{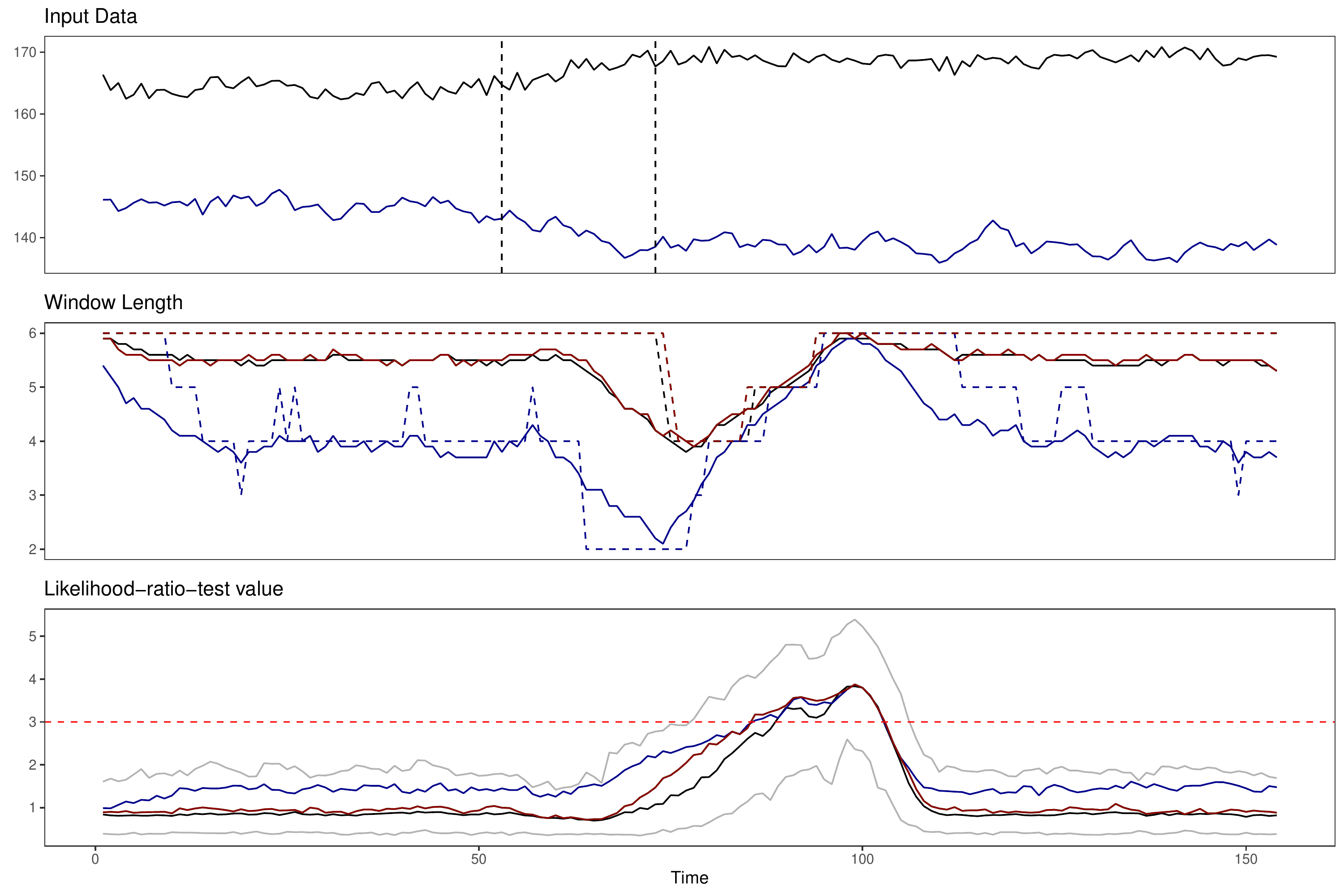}
\end{center}
\caption{Simulation robustness $d=2$ - Smooth break. \label{fig:fig9}}
\footnotesize 
\textit{Top}: Simulated 2dim VAR. Vertical lines indicate breaks. \textit{Mid}: Identified Intervals for: optimal $\rho$ (black), $\rho = 0.5$ (blue), $\rho = 0.088$ (green), optimal $\rho$ with no restrictions (red). \textit{Bottom}: Test statistics for $k+1$ with $\zeta_6 = 3.0$ (horizontal red). Bounds for 5$\%$ and 95$\%$ bounds are depicted in grey.
\end{figure}

Figure \ref{fig:fig9} illustrates the simulation results for Scenario three. It contains one break that is a linear change from $\theta_1$ to $\theta_2$: $x_1,\ldots,x_{96} \sim \theta_1, x_{97},\ldots,x_{112} \sim \theta=(\frac{16-i}{16}\theta_1 + \frac{i}{16}\theta_2)$ with $i\in [1,15], x_{113},\ldots,x_{200} \sim \theta_2$. We constructed a tiny change that is barely visible. Still, the gradual change is detected by all four specifications. However, the algorithm cannot jump down to the smallest window since the likelihood ratio is not big enough. Furthermore, while the medians follow the expected form with stairs at each window length, the mean increases more smoothly. This is because some repetitions will already result in a window length that is one higher than the current window length, which will result in a more smooth line instead of strict stairs. The scenario with the highest value of $\rho$ can return the smallest window during the break. However, it will already return smaller windows during the homogeneous part of the sample. This is because the highest value of $\rho$ will result in the smallest critical value. Hence, this scenario is best visualizing the trade-off involved in this procedure. Lower critical values will result in better detection and a higher false alarm rate. Therefore, we prefer higher critical values only to detect true breaks and to avoid false alarms, even if this means that we will not have a big break in the window length.

The likelihood ratio test values are different from previous scenarios during the break. They show a gradual reaction, slowly increasing in value. Only when the window lengths nearly recovered to their maximum again there are noticeable spikes in the test values. Despite the minimal reaction of the test statistic, particularly right after the break occurs, the algorithm still manages to detect the smooth break.

\begin{figure}[ht]
\begin{center}
\includegraphics[width=0.7\columnwidth]{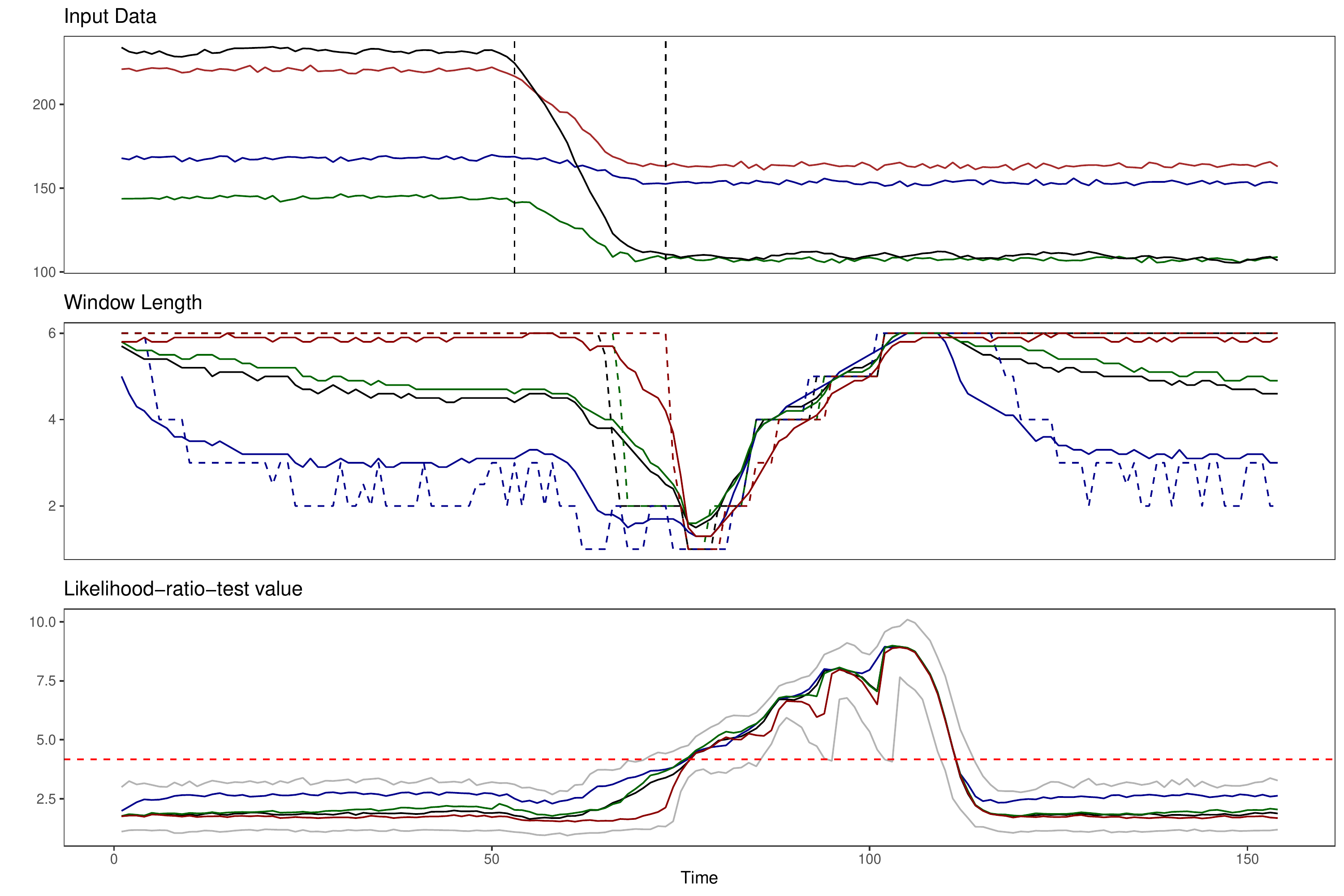}
\end{center}
\caption{Simulation robustness $d=4$ - Smooth break. \label{fig:fig10}}
\footnotesize 
\textit{Top}: Simulated 4dim VAR. Vertical lines indicate breaks. \textit{Mid}: Identified Intervals for: optimal $\rho$ (black), $\rho = 0.5$ (blue), $\rho = 0.088$ (green), optimal $\rho$ with no restrictions (red). \textit{Bottom}: Test statistics for $k+1$ with $\zeta_6 = 4.1$ (horizontal red). Bounds for 5$\%$ and 95$\%$ bounds are depicted in grey.
\end{figure}

Results for the third scenario with $d=4$ are depicted in Figure \ref{fig:fig10}. The results are even more similar to the case when $d=2$ than the other scenarios.

\hypertarget{distribution-of-test-statistics}{%
\section{\texorpdfstring{Distribution of test statistics
\label{A2}}{Distribution of test statistics }}\label{distribution-of-test-statistics}}
This appendix illustrates the distribution of test statistics for all three simulation scenarios when $d = 2$. For comparison, the sample is divided into homogeneous and non-homogeneous parts. Since the data are simulated, we know precisely which observations to use in which part. The plots are then arranged in a way that the top panel contains average test statistics for all intervals across the homogeneous part of the simulated data and the bottom panel for the non-homogeneous part. The test statistics always represent the question: "Can we extend our homogeneous interval from $k-1$ to $k$?" Therefore the number two on the x-axis represents the test when going from the shortest interval to the second one. Additionally, each plot contains a dashed horizontal red line representing the critical value for extending the homogeneous interval from the first to the second candidate interval. The critical values across intervals are very similar, so there is no need to draw them for all steps. In all top panels, the critical value is not visible since the test statistics are far below it. This is expected since the top panel shows the homogeneous samples, which should always extend to the longest candidate interval and not reject any interval length.

\subsection{Distribution of test statistics for Scenario 1: One break}\label{test-stat-scenario-1}
The first figure in this section corresponds to the results of the first scenario with just one break. The results for this scenario are shown in the main body of the paper. For the homogeneous plot, observations for $\tau \in [1,38]$ are included. This is until the Likelihood ratio test statistic in the bottom panel of Figure 2 in Chapter 5 reacts. The heterogenous plot is then based on $\tau \in [39:84]$, which includes all time points for which the test statistic deviates from the homogeneous values. As expected, some test statistics exceed the critical value only in the bottom panel. The increasing pattern in the bottom panel is due to the growing step length of the algorithm at each interval when recovering from a detected break. 

\begin{figure}[ht]
\begin{center}
\includegraphics[width=0.5\columnwidth]{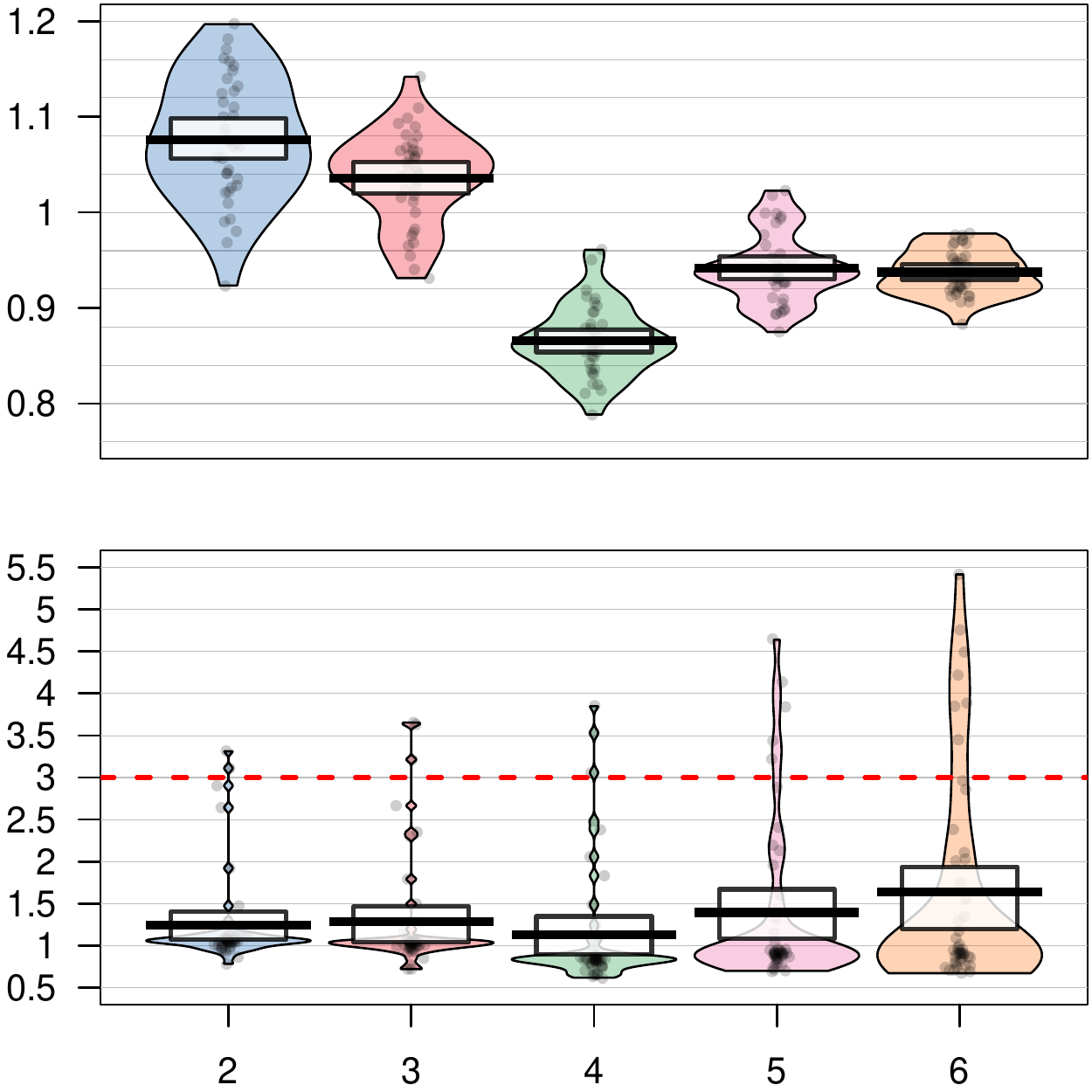}
\end{center}
\caption{Test-stats - One break. \label{fig:fig11}}
\footnotesize 
\textit{Top}: Homogeneous subsample containing 38 data points for each distribution based on $\tau \in [1,38]$. \textit{Bottom}: Non-homogeneous subsample containing 46 data points for each distribution based on $\tau \in [39:84]$. The critical value is 3.1 (horizontal red). Results are based on the median of 250 runs. The figure was created with the function pirateplot() from the package yarr in R.
\end{figure}

The clustering below the critical value occurs because it is usually just the interval covering the break going backward from each respective $\tau$ which is rejected. For example, when we are already 25 observations away from the break, the tests for intervals two, three, and four will not exceed the critical value anymore since they will not contain any breaks anymore. The few points around the critical value for each step are points where the respective window does include one or more observations from the structural gap but not enough to exceed the critical value. This behavior is by design since we want the algorithm to slowly recover after a structural break instead of jumping down and up again.

\subsection{Distribution of test statistics for Scenario 2: Two breaks}\label{test-stat-scenario-2}
\begin{figure}[ht]
\begin{center}
\includegraphics[width=0.5\columnwidth]{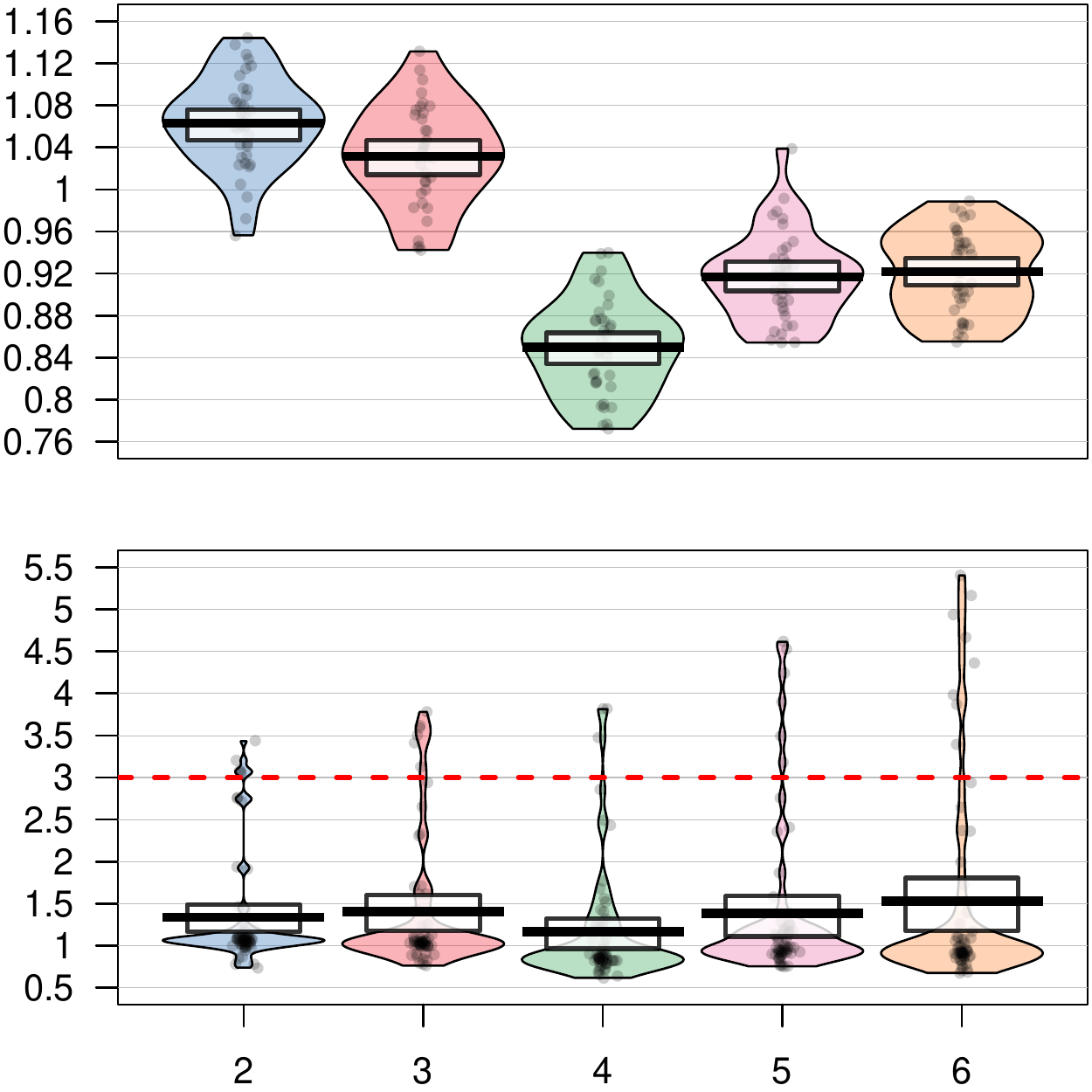}
\end{center}
\caption{Test-stats - Two breaks. \label{fig:fig12}}
\footnotesize 
\textit{Top}: Homogeneous subsample containing 38 data points for each distribution based on $\tau \in [1,38]$. \textit{Bottom}: Non-homogeneous subsample containing 62 data points for each distribution based on $\tau \in [39:100]$. The critical value is 3.1 (horizontal red). Results are based on the median of 250 runs. The figure was created with the function pirateplot() from the package yarr in R.
\end{figure}

The second test statistic figure corresponds to the second scenario, which contains two breaks within a short distance from each other and is depicted in Figure \ref{fig:fig12}. The figure looks almost identical to the one for the first scenario. The only difference is that there is a bit of a mass shift above the critical value from the highest intervals to the ones in the middle. This is because this scenario contains more data points with intervals of medium size and fewer intervals with long length since the recovery process from the first break is interrupted by the second break. In general, it makes sense that the figures are very similar since the algorithm also detected the breaks correctly when they occurred within a short distance. Therefore, our algorithm can handle structural breaks, even if they occur with high frequency in the data.

\subsection{Distribution of test statistics for Scenario 3: Smooth break}\label{test-stat-scenario-3}
\begin{figure}[ht]
\begin{center}
\includegraphics[width=0.5\columnwidth]{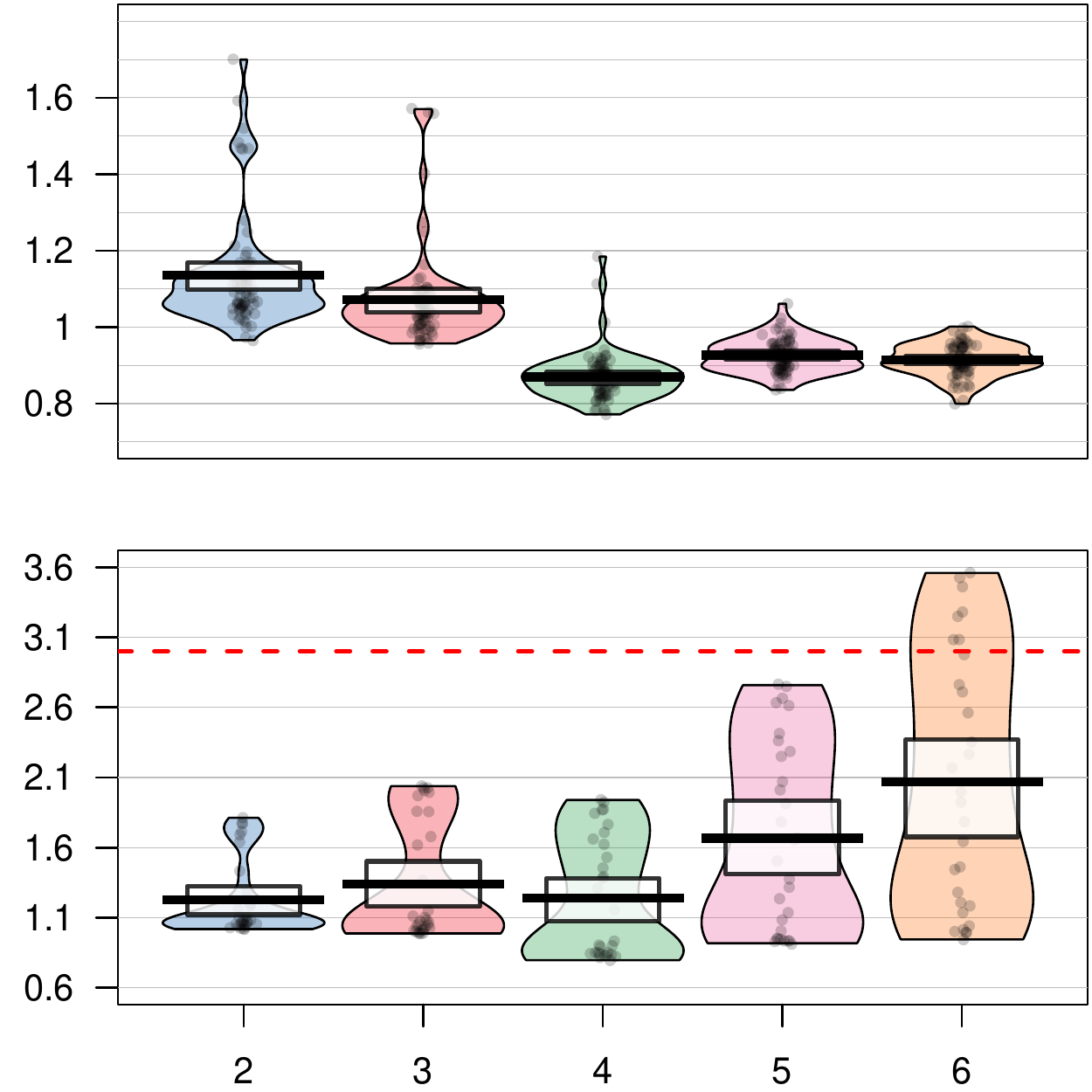}
\end{center}
\caption{Test-stats - Smooth break. \label{fig:fig13}}
\footnotesize 
\textit{Top}: Homogeneous subsample containing 70 data points for each distribution based on $\tau \in [1,70]$. \textit{Bottom}: Non-homogeneous subsample containing 30 data points for each distribution based on $\tau \in [71:100]$. The critical value is 3.1 (horizontal red). Results are based on the median of 250 runs. The figure was created with the function pirateplot() from the package yarr in R.
\end{figure}

The third test statistic figure corresponds to the simulation's third scenario, which contains a smooth break and is depicted in Figure \ref{fig:fig13}. For this scenario, even the test statistics in the bottom part are mostly smaller than the critical value. In the resulting plot of the third scenario, we can already see that the algorithm struggles to identify the break in this setting. Therefore, it is no big surprise that the test statistics, even in the non-homogeneous sample, are also minimal.

\end{document}